\documentclass[12pt,reprint,amsmath,amssymb,aps,prd,twocolumn,superscriptaddress,floatfix,nofootinbib]{revtex4-2}

\usepackage{mathtools}
\usepackage{braket}

\usepackage{bbm}

\usepackage[caption=false]{subfig}

\usepackage{tikz}
\usetikzlibrary{decorations.markings}
\usepackage[colorlinks, allcolors=blue]{hyperref}
\usepackage{booktabs}
\usepackage{multirow}

\tikzset{->-/.style={decoration={
  markings,
  mark=at position #1 with {\arrow{>}}},postaction={decorate}}}
  
\tikzset{-<-/.style={decoration={
  markings,
  mark=at position #1 with {\arrow{<}}},postaction={decorate}}}

\usepackage{csquotes}
\usepackage{amsthm}

\newtheorem{theorem}{Theorem}
\newtheorem{proposition}{Proposition}

\newcommand{\Z}{{\mathbb{Z}}}

\newcommand{\C}{{\mathbb{C}}}

\newcommand{\pqty}[1]{\left( #1 \right)}
\newcommand{\bqty}[1]{\left[ #1 \right]}
\newcommand{\abs}[1]{\left\lvert #1\right\rvert}

\newcommand{\expval}[1]{\langle #1 \rangle}

\newcommand {\dv}[3][ ]{
  \ifx #1 { }
    \frac{d #2}{d #3}
  \else
    \frac{d^{#1} #2}{d #3^{#1}}
  \fi
}

\newcommand {\pdv}[3][ ]{
  \ifx #1 { }
    \frac{\partial #2}{\partial #3}
  \else
    \frac{\partial^{#1} #2}{\partial #3^{#1}}
  \fi
}

\newcommand{\tr}{\operatorname{tr}}

\renewcommand{\Re}{\operatorname{Re}}

\newcommand{\GL}{\mathrm{GL}}
\newcommand{\Or}{\mathrm{O}}

\newcommand{\SO}{\mathrm{SO}}

\newcommand{\Sp}{\mathrm{Sp}}
\newcommand{\SU}{\mathrm{SU}}
\newcommand{\U}{\mathrm{U}}

\newcommand{\Htot}{\mathcal{H}_{\mathrm{tot}}}
\newcommand{\Hphys}{\mathcal{H}_{\mathrm{phys}}}
\newcommand{\Hhol}{\mathcal{H}_{\mathrm{hol}}}

\DeclareRobustCommand{\loongrightarrow}{%
  \DOTSB\relbar\joinrel\relbar\joinrel\relbar\joinrel\rightarrow
}

\begin{abstract}

We consider the problem of the explicit description of the gauge-invariant subspace of pure lattice gauge theories in the Hamiltonian formulation, where the gauge group is either a compact Lie group or a finite group. The latter case is particularly interesting for quantum simulation. A basis of states where configurations are grouped according to their holonomies is shown to have several advantages over other descriptions. Using this basis, we compute some properties of interest for some non-Abelian finite groups on small lattices, and in particular we examine the question of whether a certain ansatz introduced long ago is a good approximation for the ground state.

\end{abstract}

\begin{document}

\title{Almost gauge-invariant states and the ground state of Yang-Mills theory}

\newcommand{\affilBern}{\affiliation{Albert Einstein Center for Fundamental Physics, Institute for Theoretical Physics,\\
University of Bern, Sidlerstrasse 5, CH-3012 Bern, Switzerland}}

\author{A. Mariani}
\affilBern

\maketitle

\newpage
 
\section{Introduction}

The Hamiltonian formulation of lattice gauge theories \cite{KogSuss} has attracted renewed interest in recent years due to its relevance in the field of quantum simulation, i.e. the possibility of performing numerical calculations on quantum devices. On the theoretical side, quantum simulation requires a formulation of gauge theories with a finite-dimensional Hilbert space. While many proposals have been made, all of which may in principle work, here we choose to focus on the approach where one replaces the gauge group, typically a compact Lie group, with one of its finite subgroups \cite{Hasenfratz1, ZoharBurrello, Ercoetal1, Ercoetal2}. The general strategy was outlined in \cite{Hasenfratz1}: the continuum limit is approached by tuning the theory to a point in the phase diagram where the correlation length is large (i.e. a phase transition). Despite the fact that finite group theories only have a first-order transition (i.e. the correlation length becomes large but not infinite), by systematic improvement of the lattice action one may nonetheless enter the scaling regime. This has been demonstrated for the group $S(1080)$, which is the largest crystal-like subgroup of $\SU(3)$ and is therefore relevant for QCD \cite{S1080}. We also note that gauge theories with a finite gauge group are also relevant in condensed matter \cite{GiantFluctuations, DualGaugeTheory2D,DualGaugeTheory3D,CondMatDiscreteGaugeTheory2,Manjunath_2021} and quantum gravity \cite{HarlowOoguri, Toolkit}.

A peculiarity of gauge theories is that, due to the Gauss law, only a small subset of the states in their Hilbert space are physical. It is thus natural to look for a formulation purely in terms of physical degrees of freedom. Generally speaking, working in the physical subspace provides a memory advantage, in that one needs to consider fewer states; but the Hamiltonian acquires some degree of non-locality which makes operations more expensive. A further advantage of a gauge-invariant formulation is that one does not need to correct errors arising from violations of the Gauss law, which reduces memory and operational requirements. Whether this trade-off is worth making is application- and platform-dependent. For finite groups, an explicit description of the physical subspace in terms of spin networks was recently considered in \cite{MPE}.

In this work we focus on pure gauge theories with finite gauge group, but many of our results are also relevant for compact Lie groups. In Section \ref{sec:hamiltonian formulation}, we review the Hamiltonian formulation of gauge theories. In Section \ref{sec:almost}, based on a formulation introduced in the early days of lattice gauge theory \cite{Durhuus} to describe the gauge invariant content of classical observables, we consider a set of \enquote{holonomy states} which solve the Gauss law with the exception of one global constraint which is treated separately. We discuss how these states provide several advantages over the spin network formulation. Working in this basis may be useful for Hamiltonian simulations on current classical hardware as well as on current quantum devices with few qubits. We also review the issue of whether the Wilson loop observables are sufficient to describe the gauge-invariant states. Further, the issues of topology and center symmetry in this basis are discussed. Finally, in Section \ref{sec:ground state wavefunction} we perform some numerical simulations to demonstrate an application of the holonomy basis. In particular, we discuss whether a conjectured ansatz for the ground state wavefunction of Yang-Mills theory proposed in \cite{Feynman:1981ss, Greensite79, Greensite1980, fieldstrength} provides a good approximation to the exact ground state for some non-Abelian finite groups.

\section{Hamiltonian formulation}\label{sec:hamiltonian formulation}

In this section, we briefly review the Hamiltonian formulation of pure lattice gauge theories, where the gauge group is either a compact Lie group or a finite group. We also use this section to set the notation for the rest of this work.

In the lattice Hamiltonian formulation \cite{KogSuss}, time is continuous while space is given by a finite set of points, most commonly a (periodic) cubic or hypercubic lattice. In this work, however, we consider more generally the situation where space is discretized into an arbitrary finite graph (with vertices and links) which includes as a special case any type of lattice discretization. In fact, as we will see shortly, in order to define the Hilbert space and the Gauss law (which are the focus of this work) a graph structure is sufficient. Note that this excludes pathological cases such as dangling links connected to only one site. We will always assume that the graph discretizing space is connected. On the other hand, to define the magnetic Hamiltonian one requires the notion of a minimal cycle (i.e. the plaquette) for which a graph structure is not sufficient; in general, one requires the notion of a cell complex. We will not consider this here; whenever discussion of the Hamiltonian is required, we always specialize to a hypercubic lattice.

The Hilbert space is most easily described in the \enquote{group element basis}. Let $G$ be the gauge group, which is either a compact Lie group or a finite group. A classical configuration is given by an assignment of a group element to each link. In the quantum theory, states can also be found in a superposition of various group elements. Therefore we introduce a set of basis states $\ket{g}$ on each link indexed by group elements $g \in G$. These are orthonormal, i.e. $\expval{g | h} = \delta_{gh}$. The Hilbert space on each link is given by the group algebra $\C[G]$, i.e. the vector space spanned by the $\ket{g}$. For a compact Lie group, the group algebra $\C[G]$ should be replaced with the space $L^2(G)$ of square-integrable functions on the group; apart from convergence issues, its description is the same as the group algebra. Then the Hilbert space on the whole discretized space is given by a copy of $\C[G]$ on each link, i.e.
\begin{equation}
    \mathcal{H}_{\mathrm{tot}} = \bigotimes_{\mathrm{links}} \C[G] \ .
\end{equation}
We call this the \enquote{total} Hilbert space because we have not yet imposed the Gauss law. Since on each link we have $\abs{G}$ basis states, $\mathrm{dim}(\C[G]) = \abs{G}$, and therefore $\mathrm{dim} \mathcal{H}_{\mathrm{tot}} = \abs{G}^L$, where $L$ is the number of links. Basis states in the total Hilbert space are denoted as
\begin{equation}
    \ket{\{g\}} \equiv \bigotimes_{l \in \mathrm{links}} \ket{g_l} \ ,
\end{equation}
where $g_l \in G$ is an assignment of a group element to each link. In other words, the basis states in $ \mathcal{H}_{\mathrm{tot}}$ are indexed by classical configurations. Note that the graph discretizing space is assumed to have an (arbitrary) fiducial orientation. In particular, links can be traversed in either the positive or negative orientation, but if a link $l$ is traversed in the negative orientation then the group element $g_l$ is assumed to be replaced by its inverse $g_l^{-1}$.

To perform a gauge transformation, one first assigns a group element $g_x$ to each site $x$. Given such an assignment, which we denote $\{g_x\}$, the corresponding operator $\mathcal{G}(\{g_x\})$ which performs the gauge transformation acts on the basis states as
\begin{equation}
    \mathcal{G}(\{g_x\}) \ket{\{g\}} = \bigotimes_{l = \expval{xy} \in \mathrm{links}} \ket{g_x g_l g_y^{-1}}  \ ,
\end{equation}
where $l = \expval{xy}$ is the link which connects sites $x$ and $y$ (in the positive orientation). In other words, on each link one performs the transformation $\ket{g_l} \to \ket{g_x g_l g_y^{-1}}$.

It is useful to introduce the operators $L_g$ and $R_g$ which perform left- and right-translations on the group element basis states on each link. These are defined as
\begin{equation}
    \label{eq:left and right translations}
    L_g \ket{h} = \ket{gh} \ , \qquad R_g \ket{h} = \ket{h g^{-1}} \ .
\end{equation}
Both $L$ and $R$ are faithful unitary representations of the gauge group $G$, known as the left- and right-regular representations \cite{Serre, KnappLieGroups}. In terms of these operators, the gauge transformation takes the form
\begin{equation}
    \label{eq:gauge transformation}
    \mathcal{G}(\{g_x\}) = \bigotimes_{l = \expval{xy} \in \mathrm{links}} L_{g_x} R_{g_y} \ ,
\end{equation}
where it is understood that the $L$ and $R$ operators act on the Hilbert space of the corresponding link in each tensor product factor. Then one considers as belonging to the physical, gauge-invariant Hilbert space $\Hphys$ only those states $\ket{\psi}$ which are gauge-invariant \cite{KogSuss, Osborne, Tong}, i.e. 
\begin{equation}
    \mathcal{G}(\{g_x\}) \ket{\psi} = \ket{\psi} \ ,
\end{equation}
for any possible assignment $g_x$ of group elements to sites.

While our discussion of the Hilbert space is valid for both compact Lie groups and finite groups, in the rest of this work, we will only consider the Hamiltonian in the special case where the gauge group is a finite group and space is discretized as a hypercubic lattice. In this case, the Hamiltonian is given by  \cite{KogSuss, Orland, ZoharBurrello, Caspar_Wiese, HarlowOoguri, MPE}
\begin{equation}\label{eq:hamiltonian}
    H = \lambda_E \sum_{l \in \mathrm{links}} h_E + \lambda_B \sum_{\square} h_B(g_\square) \ .
\end{equation}
Here $\square$ are the plaquettes and $g_\square = g_1 g_2 g_3^{-1} g_4^{-1}$ is the associated plaquette variable, i.e. the oriented product of the four links around the plaquette. For the magnetic Hamiltonian, to match the notation of \cite{MPE} we choose
\begin{equation}
    h_B(g_{\square}) = -2 \Re \chi(g_\square) \ ,
\end{equation}
where $\chi$ is the character of a faithful representation of the gauge group $G$.

On the other hand, the electric Hamiltonian acts separately on each link, where it is given by
\begin{equation}
    \label{eq:electric hamiltonian link}
    h_E = \sum_{k \in \Gamma} (1-L_k) \ ,
\end{equation}
where $L_k$ is the left translation operator defined in  eq.\eqref{eq:left and right translations} and $\Gamma \subset G$ is a subset which doesn't contain the identity and is invariant under group conjugation and group inversion, i.e. $g \Gamma g^{-1} = \Gamma$ and $\Gamma^{-1}=\Gamma$. These two requirements ensure that $h_E$ is gauge-invariant. Note that because of these two requirements, it makes no difference whether one chooses $L_k$ or $R_k$ in eq.\eqref{eq:electric hamiltonian link}. The electric Hamiltonian on each link eq.\eqref{eq:electric hamiltonian link} is nothing but a Laplacian on the finite group \cite{MPE}. In relativistic theories, the electric and magnetic Hamiltonian are related, and one must have \cite{HarlowOoguri}
\begin{equation}\label{eq:relativistic gamma}
    \Gamma = \{g \in G,\,\, g \neq 1, \mathrm{max} \bqty{\Re \chi(g)}  \} \ ,
\end{equation}
i.e. $\Gamma$ is the subset of non-identity elements of $G$ which maximise the real part $\Re \chi$ of the character of the magnetic Hamiltonian.
As has been pointed out recently \cite{MPE}, the electric Hamiltonian on each link may be degenerate, but in this work we do not consider situations where this occurs. If this does not occur, then the electric Hamiltonian has a unique ground state on each link, given by an equal superposition over all group element states.

Finally, in Section \ref{sec:ground state wavefunction} we will numerically construct the Hamiltonian $H$ to compute some quantities of interest. We choose to measure the energy in units of $\lambda_E + \lambda_B$, so that in practice the Hamiltonian may be written as
\begin{equation}
    \label{eq:hamiltonian reduced lambda}
    H = (1-\lambda) \sum_{l \in \mathrm{links}} h_E + \lambda \sum_{\square} h_B(g_\square) \ ,
\end{equation}
where $\lambda \in [0,1]$ is a real parameter. The \enquote{strong coupling phase} where the electric Hamiltonian dominates then occurs for small $\lambda$, while for large $\lambda$ the physics is dominated by the magnetic Hamiltonian.

\section{Almost gauge-invariant states}\label{sec:almost}

As we have seen in the previous section, only a small subset of the states in the total Hilbert space are physical. In some situations, it may therefore be advantageous to work directly in the physical subspace. This reduces memory requirements, but increases the complexity of operations because the Hamiltonian generally becomes somewhat non-local. Another advantage of working in a physical basis is that one does not need to correct errors which cause violations of the Gauss law, thus contributing to reduced memory and operational requirements. Whether the advantages outweigh the disadvantages is application- and platform-dependent. 

Several approaches have been devised to describe the physical subspace of lattice gauge theories. Recently, the physical subspace of pure gauge theories whose gauge group is a finite group has been described in terms of spin-network states \cite{Baez, Burgio, MPE}. This description has several advantages, but it can be cumbersome due to the need for Clebsch-Gordan coefficients as well as the formulation in the representation basis, which together make access to operators diagonal in the group element basis quite difficult.

Since the early days of lattice gauge theory, gauge-fixing has been a popular method to reduce the number of degrees of freedom of lattice gauge theories \cite{CreutzTransferMatrix}. In this section, we will make extensive use of results in \cite{Durhuus}, where the gauge-invariant content of gauge-field configurations is described in terms of lattice holonomies based on a spanning rooted tree. This description allows the introduction of a particularly simple basis of states, which we call \enquote{holonomy states}. They are \textit{almost} gauge invariant, in the sense that they satisfy all Gauss law constraints except for one overall global transformation. We then discuss several properties of these states, including how to compute matrix elements of the Hamiltonian in this basis and their relation to a basis of Wilson loops. Interestingly, as shown in Section \ref{sec:matrix elements}, local operators in this basis involve only $\mathcal{O}(1)$ terms, against a naive expectation of $\mathcal{O}(V)$ terms where $V$ is the volume. Using this formulation, we also re-derive a formula for the dimension of the physical subspace first obtained in \cite{MPE}.

Here and throughout we use notation appropriate for finite groups. But the results of this section remain valid also for compact Lie groups, with the sole caveat that one should replace the group sums $\frac{1}{\abs{G}} \sum_{g \in G}$ with the (normalized) Haar measure $\int dg$.

The simplest way of constructing gauge-invariant states would be to start from a generic state $\ket{\psi}$ and project it onto the gauge-invariant sector. The appropriate projector can be constructed by averaging over all gauge transformations,
\begin{equation}\label{eq:gauge projector total}
    P = \frac{1}{\abs{G}^V} \sum_{\{g_x\}} \mathcal{G}(\{g_x\}) \ ,
\end{equation}
where $V$ is the number of sites. However, this method is neither explicit nor efficient. 

Alternatively, by repeatedly applying gauge transformations, the gauge field configurations split into gauge equivalence classes $X_1, X_2, \ldots$. One could then construct gauge equivalence states 
\begin{equation}\label{eq:gauge equivalence states}
    \ket{X_i} \equiv \frac{1}{\sqrt{\abs{X_i}}} \sum_{\{g\}\in X_i} \ket{ \{g\}} \ , 
\end{equation}
by equally superimposing all configurations in one gauge equivalence class. The states $\ket{X_i}$ are gauge-invariant by construction and form in fact an orthonormal basis of gauge-invariant states. One could then work directly in the physical subspace spanned by the $\ket{X_i}$. The problem is here is that the gauge equivalence classes \textit{do not} all have the same size. This can be seen already in the simple examples presented in Section \ref{sec:example}. Therefore in order to work with the states $\ket{X_i}$, one would need to know the sizes $\abs{X_i}$ of the gauge equivalence classes, but it's not clear how to compute them.

In the next section, we give two simple examples to help gain intuition for the general construction of Section \ref{sec:holonomy basis}.

\subsection{A preliminary example}\label{sec:example}

In this section, we consider two simple examples --- a $2 \times 2$ and a $2 \times 3$ square lattice with open boundaries --- in order to gain intuition for the general construction which will be presented in Section \ref{sec:holonomy basis}. Several of the issues that we will encounter in the general case are already present in these simple cases.

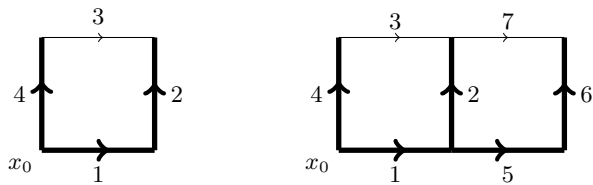
\begin{figure}
    \centering
    \begin{tikzpicture}
        \node[below left] (0,0) {$x_0$};
        \draw[->-=.6, line width=2pt] (0,0) -- (1.5,0) node[midway, below=2pt] {$1$};
        \draw[->-=.6,line width=2pt] (1.5,0) -- (1.5,1.5) node[midway, right=2pt] {$2$};
        \draw[-<-=.5,] (1.5,1.5) -- (0,1.5) node[midway, above=2pt] {$3$};
        \draw[-<-=.5,line width=2pt] (0,1.5) -- (0,0) node[midway, left=2pt] {$4$};
    \end{tikzpicture}
    \qquad \qquad
    \begin{tikzpicture}
        \node[below left] (0,0) {$x_0$};
        \draw[->-=.6,line width=2pt] (0,0) -- (1.5,0) node[midway, below=2pt] {$1$};
        \draw[->-=.6,line width=2pt] (1.5,0) -- (1.5,1.5) node[midway, right=2pt] {$2$};
        \draw[-<-=.5,] (1.5,1.5) -- (0,1.5) node[midway, above] {$3$};
        \draw[-<-=.5,line width=2pt] (0,1.5) -- (0,0) node[midway, left=2pt] {$4$};
        \draw[->-=.5,line width=2pt] (1.5,0) -- (3.0,0) node[midway, below=2pt] {$5$};
        \draw[->-=.6,line width=2pt] (3.0,0) -- (3.0,1.5) node[midway, right=2pt] {$6$};
        \draw[->-=.5] (1.5,1.5) -- (3.0,1.5) node[midway, above] {$7$};
    \end{tikzpicture}
    \caption{A $2 \times 2$ (\textit{left}) and $2 \times 3$ (\textit{right}) square lattice with open boundaries. The arrows depict the standard orientation of the links. In each case, the bold links form a possible maximal tree.}
    \label{fig:example lattices}
\end{figure}

Consider first the $2 \times 2$ lattice, shown in Fig.\ref{fig:example lattices}. This lattice has $V=4$ sites and $L=4$ links. A state in the group element basis is then of the form $\ket{g_1, g_2, g_3, g_4}$ where $g_i$ is the group element assigned to link $i$. In order to reduce the number of variables, we want to gauge-fix as many links as possible to the identity. Thus we choose a maximal tree, shown in Fig.\ref{fig:example lattices} by the bold links, and ideally we would like to set all links in the tree to the identity. This leaves out exactly one link (in this case, the one labelled $3$), and therefore the gauge-invariant content can be described by simply one variable.

This is how the story would go in the path-integral approach. However, the situation is slightly different in the Hamiltonian formulation. In fact, the \enquote{gauge-fixed} state in this case takes the form $\ket{1, 1, g_3, 1}$ and it can be easily checked that it is \textit{not} invariant under gauge transformations of the usual form eq.\eqref{eq:gauge transformation}. Instead, we construct (in this case, the only) untraced Wilson loop in this lattice, which is $h \equiv g_4 g_3 g_2^{-1} g_1^{-1}$, i.e. the plaquette. Note that we've chosen the orientation of the path defining $h$ so that it traverses link $3$ in its positive orientation. Link $3$ is precisely the one which we \textit{do not} want to fix to the identity. Also, since the Wilson loop is untraced, the base point (marked $x_0$ in Fig.\ref{fig:example lattices}) is important. As we have seen above, the gauge-invariant content on this lattice can be described by a single variable, which is the holonomy $h$. Therefore we define a state $\ket{h}$ associated to $h$ by equally superimposing all configurations with the same $h$:
\begin{equation}
    \label{eq:holonomy state 2x2}
    \ket{h} \equiv \frac{1}{\sqrt{\abs{G}}^3} \sum_{\substack{ g_1, g_2, g_3, g_4 \in G \\ g_4 g_3 g_2^{-1} g_1^{-1} = h}} \ket{g_1, g_2, g_3, g_4} \ .
\end{equation}
In his case, it is possible to check explicitly that the states $\ket{h}$ for $h \in G$ are orthonormal; in particular the normalization factor is \textit{the same} for all $h$. It is also not hard to check that the state $\ket{h}$ is invariant under all gauge-transformations except those at the base point $x_0$. This is not surprising, since $h$ is an holonomy with base point $x_0$. Therefore, we can think of $h$ as a \enquote{more} gauge-invariant version of the simple link variable $g_3$ which is the only one which we do not gauge-fix. In fact, on the gauge-fixed configuration, $h = g_3$.

But, as we said, the state $\ket{h}$ is not fully gauge-invariant. There is one remaining gauge transformation, the one at the base point $x_0$, under which $\ket{h}$ is \textit{not} invariant. This acts simply by conjugation, i.e. $\ket{h} \to \ket{g_{x_0} h g_{x_0}^{-1}}$. This means that it simply groups together the states $\ket{h}$ in the same conjugacy class. Therefore the gauge-invariant states on this lattice are labelled by the conjugacy classes $C$ of the gauge group $G$, and they are of the form
\begin{equation}
    \ket{C} \equiv \frac{1}{\sqrt{\abs{C}}} \sum_{h \in C} \ket{h} \ .
\end{equation}
These are the gauge-equivalence states of eq.\eqref{eq:gauge equivalence states}. The states $\ket{C}$ are again orthonormal, but their normalization is different for each state. In this case, it is simply given by the size of the conjugacy classes. As we will see in the other example presented in this section, computing the equivalent normalizations is in general not easy. In particular, as anticipated in the introduction to Section \ref{sec:almost}, this shows that even in this simple example the gauge-equivalence classes do not all have the same size, as the conjugacy classes of a non-Abelian group generally have different sizes. The result remains true also for compact Lie groups. In that case one cannot simply count the elements (which would be infinity); the appropriate notion which generalizes counting is volume under the Haar measure. Then the fact that different conjugacy classes have different sizes is made precise by the Weyl integration formula \cite{KnappLieGroups}.

Alternatively, the gauge-invariant states on this lattice can be described in terms of wavefunctions of the holonomy as
\begin{equation}
    \ket{\psi} = \sum_{h \in G} \psi(h) \ket{h} \ ,
\end{equation}
where the wavefunction satisfies $\psi(g h g^{-1}) = \psi(h)$. Such functions are known as \enquote{class functions} and by standard theorems \cite{Serre, KnappLieGroups} they can be expanded in a basis of irreducible characters,
\begin{equation}
    \psi(h) = \sum_i \psi_i \chi_i(h) \ ,
\end{equation}
where the $\psi_i$ are arbitrary complex coefficients and the $\chi_i$ are the irreducible characters of $G$. These are simply traces of irreps of $G$, i.e. for this lattice with only one independent holonomy, the $\chi_i(h)$ are nothing but all the traced Wilson loops in all irreducible representations of $G$. Note that it is generally true that the number of conjugacy classes of a finite group equals the number of its irreps \cite{Serre}, so that the two descriptions of the gauge-invariant states are consistent. While in this case it is possible to express the gauge-invariant states as functions of traced Wilson loops, this is not true in general. We will explore this issue further in the rest of this section as well as in Section \ref{sec:wilson loops}.

We now turn to the $2 \times 3$ square lattice shown on the right side of Fig.\ref{fig:example lattices}. In this case we have $V=6$ sites and $L=7$ links. A maximal tree like the one shown in Fig.\ref{fig:example lattices} includes $5$ links. Therefore after fully \enquote{gauge-fixing}, we are left with two group-valued variables related to the links labelled $3$ and $7$ in Fig.\ref{fig:example lattices}. But again we cannot simply gauge-fix elements to the identity, as the resulting state would not be gauge-invariant. Instead we define appropriate holonomies. In this case we have two:
\begin{equation}
    h_1 \equiv g_4 g_3 g_2^{-1} g_1^{-1} \ , \qquad h_2 \equiv g_1 g_2 g_7 g_6^{-1} g_5^{-1} g_1^{-1} \ .
\end{equation}
Importantly, we chose the two holonomies to have the \textit{same base point} $x_0$. Each holonomy goes through precisely one link which is not in the maximal tree; $h_1$ is associated with link $3$ and $h_2$ with link $7$. Now we define holonomy states
\begin{equation}
    \ket{h_1, h_2} \equiv \frac{1}{\sqrt{\abs{G}}^5} \sum_{\{g\} \in [h_1, h_2]} \ket{\{g\}} \ ,
\end{equation}
by equally superimposing all configurations with the same holonomies. Again it is tedious but not hard to check that the states $\ket{h_1, h_2}$ are orthonormal. In particular, they again all have the same normalization, which is the same as saying that all classes of states with the same holonomies contain an equal number of states. We will show that this statement is true in general in Section \ref{sec:holonomy basis}.

The holonomy states $\ket{h_1, h_2}$ are invariant by construction under all gauge transformations except the one at the base point $x_0$. Under this gauge transformation, they transform by simultaneous conjugation, i.e. as $\ket{h_1, h_2} \to \ket{g_{x_0} h_1 g_{x_0}^{-1}, g_{x_0} h_2 g_{x_0}^{-1}}$. Unlike the previous case with only one holonomy, this condition admits no simple mathematical characterization \footnote{In particular, it is not conjugation in the group $G \times G$, which would allow the two holonomies to be conjugated by different group elements.}. One could still form gauge-equivalence states by superimposing holonomy states related by simultaneous conjugation, but it is not clear how (apart from brute force) to compute the size of the simultaneous conjugation classes and therefore the normalization of the gauge-equivalence states.

Nonetheless, we can simply describe the gauge-invariant Hilbert space as the space of states of the form
\begin{equation}
    \ket{\psi} = \sum_{h_1, h_2 \in G} \psi(h_1, h_2) \ket{h_1, h_2} \ ,
\end{equation}
where the wavefunction $\psi$ satisfies $\psi(g h_1 g^{-1},g h_2 g^{-1}) = \psi(h_1, h_2)$ for all $g \in G$. Unfortunately, again, functions which are invariant under simultaneous conjugation admit no simple characterization. Given a character $\chi$ of $G$, all the traced Wilson loops $\chi(h_1)$, $\chi(h_2)$, $\chi(h_1 h_2)$ (or more generally $\chi(h)$ where $h$ is an arbitrary product of $h_1$ and $h_2$) are gauge-invariant, but there is no guarantee that they \textit{span} the space of gauge-invariant functions. In fact, as we will see in Section \ref{sec:wilson loops}, this is not true in general.

In the next section, we generalize the construction of this chapter to the case where space is discretized as an arbitrary graph. In particular, we construct holonomy states in full generality. In Section \ref{sec:wilson loops} we discuss under what conditions Wilson loops are sufficient to describe the gauge-invariant functions. Then in Section \ref{sec:matrix elements} we show how to construct matrix elements of the Hamiltonian in the holonomy basis.

\subsection{The holonomy basis}\label{sec:holonomy basis}

In this section, we generalize the construction exemplified in the previous section to arbitrary graph geometries. We describe a basis of \enquote{holonomy states} which can be implemented without requiring knowledge of any coefficients or normalizations. They satisfy all Gauss law constraints except one, which is then treated separately.

In what follows, we use the word \enquote{path} in a loose sense. In particular, we say that a path $\gamma$ between two sites $x$ and $y$ in a graph $\Lambda$ is a sequence of sites $x_0, x_1, \ldots, x_n \in \Lambda$ such that $\expval{x_i, x_{i+1}}$ is a link in $\Lambda$ for each $i$ and $x_0=x$, $x_n = y$. The links can be traversed in either the positive or negative orientations indifferently. Then a path $\gamma$ is also naturally associated with an ordered set of links in the graph, i.e. the links $\expval{x_i, x_{i+1}}$.

\begin{figure*}
    \centering
    \begin{tikzpicture}
        \draw[step=1.0,black] (1,1) grid (4,3);
        \node[below left] at (1,1) {$x_0$};
        \draw[line width=2pt] (1,1) -- (4,1);
        \draw[line width=2pt] (1,1) -- (1,3);
        \draw[line width=2pt] (2,1) -- (2,3);
        \draw[line width=2pt] (3,1) -- (3,3);
        \draw[line width=2pt] (4,1) -- (4,3);
    \end{tikzpicture}
    \qquad\qquad\qquad\qquad
    \begin{tikzpicture}
        \draw[->-=0.5] (1,1) -- (2,1);
        \draw[->-=0.5] (1,1) -- (1,2);
        \draw[white] (1, 0) -- (2, 0);
    \end{tikzpicture}
    \qquad\qquad\qquad
    \begin{tikzpicture}
        \draw[step=1.0,black] (1,1) grid (4,3);
        \node[below left] at (1,1) {$x_0$};
        \draw[line width=2pt] (1,1) -- (4,1);
        \draw[line width=2pt] (1,1) -- (1,3);
        \draw[line width=2pt] (2,1) -- (2,3);
        \draw[line width=2pt] (3,1) -- (3,3);
        \draw[line width=2pt] (4,1) -- (4,3);
        \draw[->-=0.5, dotted] (1,1.1) -- (1.9,1.1);
        \draw[->-=0.5, dotted] (1.9,1.1) -- (1.9,2.1);
        \draw[->-=0.5, dotted] (1.9,2.1) -- (3.1,2.1);
        \draw[->-=0.5, dotted] (3.1,2.1) -- (3.1,0.9);
        \draw[->-=0.5, dotted] (3.1,0.9) -- (1.0,0.9);
    \end{tikzpicture}
    \caption{Illustration of the construction of the holonomies on a $3 \times 4$ square lattice with open boundaries. The reference orientation of the links is shown in the center figure. On the left, the lattice together with a maximal tree (the bold links) and root $x_0$. On the right, a path making up an holonomy according to the construction described in the text.}
    \label{fig:holonomy example}
\end{figure*}

Given a gauge field configuration, we can associate to each path $\gamma$ given by links $l_1, l_2, \ldots, l_n$ a Wilson line $h_\gamma \in G$ defined as
\begin{equation}
    h_\gamma = \prod_{i=1}^n g_{l_i}^{\sigma(l_i, \gamma)} \ ,
\end{equation}
where $\sigma(l_i, \gamma) = \pm 1$ depending on whether $\gamma$ traverses $l_i$ in the positive or negative orientation. Note that, importantly, no trace is taken. If the path $\gamma$ starts and ends at the same point, we refer to $h_\gamma$ as a \enquote{holonomy}. Since the holonomies are defined without a trace, the base point is important.

Now we follow closely the procedure described in \cite{Durhuus}. Starting from the graph $\Lambda$ which discretizes space, pick a maximal spanning tree $T$, i.e. a subgraph of $\Lambda$ which includes all vertices of $\Lambda$ but has no closed loops. (A connected graph always has a spanning tree \cite{Foulds1992}) Pick an arbitrary special site $x_0$, which we call the \textit{root}. Examples of spanning trees are shown in Fig.s \ref{fig:example lattices} and \ref{fig:holonomy example}. By construction $T$ has $V$ vertices (the same as $\Lambda$) and, by the general properties of tree graphs, $V-1$ links \cite{Foulds1992}. Therefore $\Lambda$ has $L$ links in total, of which $V-1$ are in $T$ and $M \equiv L-V+1$ outside $T$. Note also that between any two sites $x$ and $y$ in $\Lambda$ there is a unique path (i.e. a unique set of links) lying entirely within $T$ \cite{Foulds1992}. 
The construction is illustrated in Fig.~\ref{fig:holonomy example}. Given a gauge field configuration, construct a specific set of holonomies associated to it, which we call the \textit{fundamental holonomies}, as follows:
\begin{enumerate}
    \item For each link $\widetilde{l}=\expval{xy} \in \Lambda \setminus T$, i.e. \textit{outside} $T$, construct the closed path $\gamma$ which starts from the root $x_0$, follows the unique path in $T$ from $x_0$ to $x$, follows  $\widetilde{l}$ to reach $y$ and then follows the unique path in $T$ from $y$ back to $x_0$.
    \item Construct the holonomy associated to $\gamma$.
    \item Note that with this procedure we have constructed $M=L-V+1$ independent holonomies, each associated to a link outside the tree. We call these the \textit{fundamental holonomies}. Any other holonomy can be obtained by a finite product of fundamental holonomies.
\end{enumerate}

In particular, note that the fundamental holonomies are invariant under all gauge transformations which equal the identity at the root, i.e. $g_{x_0}=1$. They are \textit{not} invariant under gauge transformations at the root. For ease of reference we call \textit{internal} gauge transformations those which equal identity at the root, i.e. $g_{x_0}=1$. We call the remaining gauge transformation, the one based at the root $x_0$, the \textit{external} gauge transformation. Then we say that the fundamental holonomies are invariant under the internal gauge transformations, which form a group $\mathcal{G}_\mathrm{int}$; on the other hand, under the external gauge transformation all holonomies transform in the same way, i.e. $h_\gamma \to g_{x_0} h_\gamma g_{x_0}^{-1}$. The external gauge transformations also form a group which we call $\mathcal{G}_\mathrm{ext}$.

Among all gauge field configurations with the same set of $M$ holonomies $\{h\}$, there is one which is particularly simple. It is the one where all links in the maximal tree $T$ are set to the identity, and we refer to it as the \enquote{gauge-fixed} configuration. Formally we denote it as
\begin{equation}
    \ket{\{h\}_{\mathrm{GF}}} \equiv \bigotimes_{l \in \mathrm{links}} \ket{g_l} \qquad g_l = \begin{cases} h_l & l \in \Lambda \setminus T \\ 1 & l \in T\end{cases} \ ,
\end{equation}
where here $h_l$ is the holonomy associated to link $l$ outside the tree. It is clear that the gauge field configuration $\ket{\{h\}_{\mathrm{GF}}}$ has fundamental holonomies $\{h\}$.

As is clear from the construction, knowledge of the fundamental holonomies and of the gauge field on the tree is sufficient to reconstruct the gauge field configuration. Since also general (non-fundamental) holonomies are invariant under internal gauge transformations, this also implies that the fundamental holonomies with root $x_0$ are sufficient to reconstruct all holonomies with the same root $x_0$. In fact, since they can be equivalently evaluated on the gauge-fixed configuration, all holonomies with root $x_0$ are given by products of the fundamental holonomies. Moreover, any holonomy with arbitrary base point is given, up to conjugation, by a product of fundamental holonomies; if $\gamma$ is a closed path with base point $x \neq x_0$ and $\omega$ is the path from $x_0 \to x$ in the spanning tree, then $\omega \circ \gamma \circ \omega^{-1}$ is another closed path, but with base point $x$. An example of this occurs in Fig.\ref{fig:holonomy example}. In what follows, we will often refer to the fundamental holonomies simply as the \enquote{holonomies}.

Now we show an important result for what follows:

\begin{proposition}\label{prop:equivalence to gauge fixed}
Let $\ket{\{g\}}$ be a gauge field configuration with fundamental holonomies $\{h\}$. Then $\ket{\{g\}}$ is internally gauge-equivalent to the gauge-fixed configuration, i.e. there is an internal gauge transformation $\mathcal{G}$ such that $\ket{\{g\}} = \mathcal{G} \ket{\{h\}_{\mathrm{GF}}}$.
\end{proposition} 

Variations on this result have certainly appeared before in the literature, see for example \cite{Cui} for a similar statement and proof.

\begin{proof}
We will exhibit a sequence of internal gauge transformations which gauge-fixes to the identity the gauge field in the tree. Note that a product of internal gauge transformations is also an internal gauge transformation, and that internal gauge transformations preserve the holonomies. Therefore the end result of this procedure is the gauge-fixed configuration.

Consider the spanning tree $T$. Call \textit{active} the sites at which we have not yet fixed the gauge transformation, and \textit{inactive} otherwise. If the gauge field $g_l$ at link $l$ has been set to the identity by a gauge transformation, we mark it as \textit{frozen}. Initially, the root site is inactive and all other sites are active. A \textit{leaf} of the tree $T$ is a site which is connected only to one other site. Every tree has at least one leaf \cite{Foulds1992}. Then pick a leaf $x$ and consider the path from the root $x_0$ to $x$ (this is unique \cite{Foulds1992}), which takes the form $x_0, x_1, x_2, \ldots, x_k$ with $x_k=x$. Then use the gauge freedom at $x_1$ to set the gauge field on link $l = \expval{x_0 x_1}$ to the identity; then mark $x_1$ as inactive and $l$ as frozen. Note that since both sites connected to $l$ are inactive, its value cannot be changed by the following gauge transformations. Then continue to use the gauge freedom at site $x_i$ to fix the gauge field on link $\expval{x_{i-1}x_i}$. Thus all the links in the path between $x_0$ and $x_k$ are frozen and all the sites in the path are inactive. Now among inactive sites, choose one (say $y_0$) which is connected to a link which is not frozen. If it does not exist, then all links are frozen and the algorithm terminates. Otherwise, consider the subgraph formed by $y_0$ and all sites and edges reachable via non-frozen links. Since this is a subgraph of $T$, it is again a tree and therefore will have a leaf. Thus we can repeat the previous algorithm whereby we consider the path from $y_0$ to the leaf and fix the gauge field along the path. Then one again picks another inactive site recursively until all links in the tree are frozen.
\end{proof}

Using the previous proposition, we can show the following:

\begin{proposition} \label{prop:configurations same holonomies}
Two gauge field configuration states $\ket{\{g\}}$ and $\ket{\{g'\}}$ have the same fundamental holonomies if and only if they are related by an internal gauge transformation.
\end{proposition}

\begin{proof}
One direction is obvious. We know that internal gauge transformations do not change the holonomies; therefore if $\ket{\{g\}}$ and $\ket{\{g'\}}$ are related by an internal gauge transformation, then they have the same holonomies. For the other direction, we make use of Proposition \ref{prop:equivalence to gauge fixed} according to which both $\ket{\{g\}}$ and $\ket{\{g'\}}$ are related to the gauge-fixed configuration via an internal gauge transformation. The result then follows because internal gauge transformations form a group.
\end{proof}

All the results until now deal with gauge field configurations and are therefore well-known. Now we deal more specifically with the consequences for the physical Hilbert space.

We define a \textit{holonomy class} to be the equivalence class of gauge field configurations with the same fundamental holonomies. We have said before that gauge equivalence classes do not all have the same number of elements. The remarkable fact about \textit{holonomy classes} is that they \textit{do} all have the same number of elements. Before proving this fact, we show a refinement of the previous propositions,

\begin{proposition} \label{prop:one-to-one corr}
Within each holonomy class, configuration states are in one-to-one correspondence with internal gauge transformations.
\end{proposition}

\begin{proof}
Given a configuration state $\ket{\{g\}}$ with holonomies $\{h\}$, from Proposition \ref{prop:equivalence to gauge fixed} we know that there is an internal gauge transformation $\mathcal{G}$ such that $\ket{\{g\}} = \mathcal{G} \ket{\{h\}_{\mathrm{GF}}}$. Then we define a map from configurations to internal gauge transformations by $\ket{\{g\}} \to \mathcal{G}$ and we have to show that it is one-to-one.

First of all, we have to show that the map is well-defined. In particular, we have to show that if we have two gauge transformations $\mathcal{G}$ and $\mathcal{G}'$ such that
\begin{align}\label{eq:possible double}
    \ket{\{g\}} &= \mathcal{G} \ket{\{h\}_{\mathrm{GF}}} \ , \\
    \ket{\{g\}} &= \mathcal{G}' \ket{\{h\}_{\mathrm{GF}}} \ ,
\end{align}
then $\mathcal{G} = \mathcal{G}'$. In fact uniqueness follows from the proof of Prop. \ref{prop:equivalence to gauge fixed} because gauge transformations on different sites commute and paths in a tree are unique. But we prove this result again here in a simpler manner. Since internal gauge transformations form a group, they have inverses. Therefore from eq.\eqref{eq:possible double} we have 
\begin{equation}
    \ket{\{h\}_{\mathrm{GF}}} = \mathcal{G}^{-1} \mathcal{G}'   \ket{\{h\}_{\mathrm{GF}}}  \ .
\end{equation}
Therefore the gauge-fixed configuration is fixed by the internal gauge transformation $\mathcal{G}' \mathcal{G}^{-1}$. But in the gauge-fixed configuration, each link in the tree is set to the identity. For a link $l = \expval{xy}$ in the tree, $\mathcal{G}' \mathcal{G}^{-1}$ acts as $\ket{1} \to \ket{g_x g_y^{-1}} = \ket{1}$ for some group elements $g_x, g_y \in G$. Therefore $g_x = g_y$. But since the tree is maximal, i.e. it includes all vertices, this then implies that $\mathcal{G}' \mathcal{G}^{-1}$ is given by the same group element on \textit{all} sites. But since it is an internal gauge transformation, it is equal to the identity on the root $x_0$; therefore it must be given by the identity on \textit{all} sites, i.e. it is the identity transformation. Therefore $\mathcal{G}' = \mathcal{G}$.

It remains to show that the map is one-to-one. But this can be done by exhibiting an explicit inverse, which is simply given by the map $\mathcal{G} \to \ket{\{g\}} \equiv \mathcal{G} \ket{\{h\}_{\mathrm{GF}}}$.
\end{proof}

Incidentally, this proves that the group $\mathcal{G}_{\mathrm{int}}$ of internal gauge transformations is really isomorphic to $G^{V-1}$. On the other hand, the group of external gauge transformations $\mathcal{G}_{\mathrm{ext}}$ is \textit{not} isomorphic to $G$. In fact it is clear that if $g$ is in the center $Z(G)$ of $G$, then the corresponding external gauge transformation $\mathcal{G}(g)$ is simply the identity. Therefore $\mathcal{G}_{\mathrm{ext}} \cong G / Z(G)$. Similarly, the full group of gauge transformations is isomorphic to $G^V / Z(G)$ where $Z(G)$ is the \enquote{diagonal} subgroup of constant transformations equal to a central element (which act as the identity).

We can now show that all holonomy classes have the same number of states:

\begin{proposition}
    Each holonomy class contains $\abs{G}^{V-1}$ configurations.
\end{proposition}

\begin{proof}
Suppose that we want to construct a configuration with prescribed fundamental holonomies. These are always defined with respect to a spanning tree, say $T$. Following Propositions \ref{prop:equivalence to gauge fixed} and \ref{prop:configurations same holonomies}, we're free to set the gauge field configuration on each link in the tree arbitrarily. Then for each link not in $T$, we assign to it the unique group element which fixes its fundamental holonomy to the prescribed one. Hence we're free to set all and only the links on the tree $T$. Since there are $V-1$ such links, we can construct $\abs{G}^{V-1}$ states with the same fundamental holonomies.
\end{proof}

This statement can also be proved from an alternative point of view:

\begin{proof}
    From Proposition \ref{prop:one-to-one corr} within each holonomy class configurations are in one-to-one correspondence with internal gauge transformations. Therefore it suffices to count the number of internal gauge transformations. Since we are free to choose $\abs{G}$ elements for $V-1$ sites (i.e. all sites except the root), then we have $\abs{G}^{V-1}$ internal gauge transformations and therefore the same number of configurations in each holonomy class. 
\end{proof}

This means, in particular, that we can define the \enquote{holonomy class states}
\begin{equation}
    \ket{\{h\}} \equiv \ket{h_1, h_2, \ldots, h_M} \equiv \tfrac{1}{\sqrt{\abs{G}}^{V-1}} \sum_{\{g\} \in [h]} \ket{\{g\}} \ ,
\end{equation}
by an equal superposition of all configuration states with the same holonomies. Since all holonomy classes have the same size, the normalization for these states is known and in fact the states $\ket{\{h\}}$ are orthonormal,
\begin{equation}
    \expval{\{h'\}| \{h\} } = \prod_{i=1}^{M} \delta(h_i', h_i) \ .
\end{equation}
 We call the Hilbert space spanned by the holonomy states $\Hhol$. Since each holonomy can take $\abs{G}$ values and we have $M=L-V+1$ holonomies, then $\mathrm{dim} \Hhol = \abs{G}^{L-V+1}$. In particular, $\Hhol$ admits a remarkably simple description as a tensor product space,
\begin{equation}
    \Hhol = \C[G]^{\otimes M} \ .
\end{equation}
In other words, $\Hhol$ is simply the Hilbert space spanned by states of the form $\ket{h_1, h_2, \ldots h_M}$ where each holonomy $h_i \in G$ can take any value in $G$ without restrictions. Therefore the holonomy states satisfy almost all gauge constraints, thus reducing the size of the Hilbert space by an exponential factor, without requiring the computation of normalizations or Clebsch-Gordan coefficients. One is simply left with one overall global constraint, i.e. the external gauge transformation. Note in particular that the description of the holonomy space is entirely independent of the embedding of the holonomies into any particular graph, only \textit{operators} depend on the embedding. 

The holonomy states are generally not gauge-invariant because of the external gauge transformations $\mathcal{G}_{\mathrm{ext}}$. In fact a transformation $\mathcal{G}(g) \in \mathcal{G}_{\mathrm{ext}}$ acts in holonomy space as
\begin{equation}
    \mathcal{G}(g) \ket{h_1, h_2, \ldots, h_M} = \ket{g h_1 g^{-1}, g h_2 g^{-1}, \ldots, g h_M g^{-1}} \ .
\end{equation}
In holonomy space, the external gauge transformation takes the role of a global symmetry which acts by simultaneous conjugation of all holonomies. Therefore the Hilbert space of holonomies is an intermediate space between the total Hilbert space, and the physical Hilbert space:
\begin{equation}
    \Htot \overset{\mathcal{G}_{\mathrm{int}}}{\loongrightarrow} \Hhol \overset{\mathcal{G}_{\mathrm{ext}}}{\loongrightarrow} \Hphys \ .
\end{equation}
The projector from $\Hhol$ to $\Hphys$ is then constructed by averaging over external gauge transformations, i.e. it is given by the operator
\begin{equation}
    P \ket{h_1, \cdots, h_M} = \frac{1}{\abs{G}} \sum_{g \in G} \ket{g h_1 g^{-1}, \ldots, g h_M g^{-1}} \ .
\end{equation}
In terms of the $L$ and $R$ operators defined in Section \ref{sec:hamiltonian formulation} (now acting on the Hilbert space of each holonomy), it is given by
\begin{equation}\label{eq:gauge projector holonomies}
    P = \frac{1}{\abs{G}} \sum_{g \in G} (L_g R_g)^{\otimes M} \ .
\end{equation}
It is not hard to show that $P$ is indeed a projector. We can then use it to re-derive a formula for the dimension of the physical subspace first obtained in \cite{MPE}:

\begin{theorem}\label{theo:Hphys size}
    If $C$ are the conjugacy classes of $G$, and $\abs{C}$ their sizes, then on a graph with $V$ vertices and $L$ edges one has $\mathrm{dim} \Hphys = \sum_C \pqty{\frac{\abs{G}}{\abs{C}}}^{L-V}$.
\end{theorem}

\begin{proof}
    As we have seen, $P$ projects onto the gauge-invariant sector. As usual, the dimension of the image of the projector (in this case, the gauge-invariant subspace) equals the trace of the projector. Therefore
    \begin{equation}
        \mathrm{dim} \Hphys = \tr{P} = \frac{1}{\abs{G}} \sum_{g \in G} \bqty{\tr(L_g R_g)}^M \ .
    \end{equation}
    By the Peter-Weyl decomposition \cite{KnappLieGroups, marianithesis}, we find 
    \begin{equation}
        L_g R_g = \bigoplus_j \rho_j(g)^* \otimes \rho_j(g) \ ,
    \end{equation}
    where $\rho_j$ are the irreps of $G$. Then taking traces, one finds
    \begin{equation}
        \tr(L_g R_g) = \sum_j \chi_j(g)^* \chi_j(g) = \frac{\abs{G}}{\abs{C(g)}}
    \end{equation}
    where we used one of the orthogonality relations of characters \cite{Serre} and $C(g)$ is the conjugacy class of $g$. Therefore we find 
    \begin{align}
        \mathrm{dim} \Hphys &= \tr{P} = \frac{1}{\abs{G}} \sum_{g \in G } \frac{\abs{G}^{L-V+1}}{\abs{C(g)}^{L-V+1}} =\\
        &=\sum_C \sum_{g \in C } \frac{\abs{G}^{L-V}}{\abs{C}^{L-V+1}} = \sum_C \frac{\abs{G}^{L-V}}{\abs{C}^{L-V}}
    \end{align}
    which proves the result.
\end{proof}

In fact this statement could also have been proven by computing the trace of the projector from the total Hilbert space, eq.\eqref{eq:gauge projector total}, by combining the argument presented here with the one in \cite{MPE}.

Note that in an Abelian group, all elements commute and therefore $g h_i g^{-1}=h_i$ so that the external gauge transformation does nothing. In this case, therefore the space of holonomies coincides with the physical Hilbert space, $\Hhol \cong \Hphys$. 

The statement that all holonomy classes have the same size may also be understood as the statement that the transformation from configurations to holonomies has constant Jacobian. Consider in fact a gauge-invariant function of configurations $f(\{g\})$, i.e. $f: G^L \to \C$. This may also be thought of as the wavefunction of a gauge-invariant state. Since $f$ is gauge-invariant, it depends only on the holonomies of $\{g\}$. Therefore we can define a function of the holonomies $F: G^M \to \C$ via $F(\{h\}) \equiv f(\{h\}_{\mathrm{GF}})$. Then we have the change of variable formula,
\begin{equation}
    \label{eq:change of variable formula}
    \frac{1}{\abs{G}^L} \sum_{\{g\}} f(\{g\}) =\frac{1}{\abs{G}^M} \sum_{\{h\}} F(\{h\}) \ .
\end{equation}
Here $\frac{1}{\abs{G}} \sum_{g \in G}$ should be thought of as the normalized Haar measure on the finite group; the same formula also holds for compact Lie groups by replacing this with their Haar measure.

It is also interesting to compare the dimension of the two spaces. One has
\begin{equation}
    \frac{\mathrm{dim} \Hphys}{\mathrm{dim} \Hhol} = \frac{1}{\abs{G}} \sum_C \frac{1}{\abs{C}^{L-V}} \gtrsim \frac{1}{\abs{G}} \ .
\end{equation}
Therefore, in going to the physical subspace one gains at most a factor of $\approx \abs{G}$ in terms of \enquote{memory} compared to the holonomy space, paying a price in terms of having to store the Clebsch-Gordan coefficients which express the gauge-invariant states in terms of the holonomy states.

\subsection{Wilson loops as a basis} \label{sec:wilson loops}

For $\SU(N)$ gauge theories, it is standard to expand gauge-invariant functions in terms of traced Wilson loops. These are generally overcomplete, but at least for $\SU(2)$ it was possible to find a subset of loops which form a proper basis \cite{Watson_1994}. As anticipated in Section \ref{sec:example}, in general the traced Wilson loops do not even span the space of gauge-invariant functions.

In this section, we review the issue of whether a basis of gauge-invariant states may be found in terms of traced Wilson loops for a general finite or compact Lie gauge group $G$. This issue is closely related to the discussion of this section and the results discussed here have been obtained by several authors \cite{Durhuus, Sengupta, Levy, Cui}. From the previous discussion, in terms of the holonomy basis, gauge-invariant states take the form
\begin{equation}
    \ket{\psi} = \sum_{\{h\}} \psi(h_1, h_2, \ldots, h_M) \ket{h_1, h_2, \ldots, h_M} \ ,
\end{equation}
where the wavefunction in the holonomy basis satisfies
\begin{equation}
    \label{eq:gauge-invariant wavefunction condition}
     \psi(g h_1 g^{-1},g h_2 g^{-1}, \ldots,g h_M g^{-1}) = \psi(h_1, h_2, \ldots, h_M) \ ,
\end{equation}
for all $g \in G$. As we saw in Section \ref{sec:example}, for $M=1$ this is simply a class function on $G$ (a basis of which is given by the characters of $G$), but for $M > 1$ this condition admits no such simple mathematical description. One may nonetheless attempt to construct a basis of such scalar functions, which would then also provide a basis of gauge-invariant states. A simple observation is that a set of functions which satisfies eq.\eqref{eq:gauge-invariant wavefunction condition} are of the form $f(h)$ where $f$ is a class function on $G$ and $h$ is a product of fundamental holonomies, i.e.
\begin{equation}
    h = h_{i_1} h_{i_2} h_{i_3} \cdots \ ,
\end{equation}
where each fundamental holonomy appears an arbitrary number of times. But since the irreducible group characters form a basis of class functions, it is equivalent to consider a basis of scalar functions of the form $\chi(h)$, where $\chi$ is an irreducible character and $h$ a product of holonomies. These are precisely all the possible traced Wilson loops in all the irreducible representations of $G$.

One might then attempt to prove that all functions satisfying eq.\eqref{eq:gauge-invariant wavefunction condition} may be expanded in a basis of traced Wilson loops; but this turns out not to be possible. This is not only because such a basis would be overcomplete, but more importantly because \textit{the traced Wilson loops do not necessarily span the space of functions satisfying eq.}\eqref{eq:gauge-invariant wavefunction condition}. In fact for some gauge groups $G$ it is possible to construct \textit{orthogonal} gauge-invariant states with the \textit{same} Wilson loops \cite{Cui}. 

Nonetheless, the traced Wilson loops \textit{do} span the space of gauge-invariant functions in some cases of interest (while still being overcomplete). For example if the group $G$ is Abelian, then the condition eq.\eqref{eq:gauge-invariant wavefunction condition} is trivial. For $\Z_N$ and $\U(1)$ therefore we obtain the even stronger result that Wilson loops in any one faithful irreducible representation are sufficient to determine the gauge-invariant content of configurations. But this result is in a sense not very interesting, as the above description in terms of holonomy states is already fully sufficient to describe the gauge-invariant states. 

More generally, note that if $G$ is either finite or compact Lie and $\chi(g)=\chi(g')$ for all irreducible characters $\chi$, then $g$ and $g'$ are conjugate \cite{Serre, KnappLieGroups}. Suppose that one has two states with holonomies $\{h\}$ and $\{h'\}$. Then one can consider all possible independent products of holonomies, which we call $\widetilde{h}_\alpha$ and $\widetilde{h}'_\alpha$ indexed by $\alpha$. If the Wilson loops form a basis of gauge-invariant states then they must agree on $\{h\}$ and $\{h'\}$ if and only if the two are gauge-equivalent. From the previous result we know that if $\chi(\widetilde{h}_\alpha) = \chi(\widetilde{h}'_\alpha)$ for all $\chi$ (i.e. the Wilson loops agree), then $\widetilde{h}_\alpha$ and $\widetilde{h}'_\alpha$ are conjugate. But $\{h\}$ and $\{h'\}$ are gauge-equivalent if and only if $h_i' = g h_i g^{-1}$ for some $g$, which must be the \textit{same} for all holonomies; therefore it is not sufficient that $\widetilde{h}_\alpha$ and $\widetilde{h}'_\alpha$ are conjugate for all $\alpha$, but they must be conjugate \textit{by the same group element for all } $\alpha$. In order to guarantee this, one must exclude the presence of certain group homomorphisms. In fact suppose that $\phi$ is a class-preserving outer automorphism \cite{Sah1968, brooksbank2013groups, Durhuus, Cui}, i.e. $\phi: G \to G$ is an invertible group homomorphism such that $g$ and $\phi(g)$ are always in the same conjugacy class, but $\phi(g)$ is \textit{not} of the form $\phi(g) = k g k^{-1}$ for some $k \in G$. Then the configurations $\{g\}$ and $\{\phi(g)\}$ have holonomies $\{h\}$ and $\{\phi(h)\}$ and identical Wilson loops, $\chi(h) = \chi(\phi(h))$ on account of $\phi$ preserving conjugacy classes. Yet the configurations $\{g\}$ and $\{\phi(g)\}$ are not gauge-equivalent, since $\phi(g) \not\equiv k g k^{-1}$ for fixed $k \in G$ and therefore their holonomies are not related by an external gauge transformation.  But note that it is \textit{not} sufficient to exclude automorphisms $\phi : G \to G$, because one may consider gauge field configurations lying entirely within a subgroup of $G$. More generally, one must prove that for any two (closed) subgroups $H_1$ and $H_2$ of $G$, an isomorphism $\phi: H_1 \to H_2$ such that $g$ and $\phi(g)$ are conjugate in $G$, must be of the form $\phi(g) \equiv k g k^{-1}$ for all $g \in H_1$ and some $k \in G$ \cite{Durhuus}.

Putting together the results of several authors \cite{Durhuus, Sengupta, Levy}, one finds that Wilson loops determine gauge-invariant states if the gauge group $G$ is a finite direct product of Abelian groups together with any finite number of copies of the groups $\U(N)$, $\SU(N)$, $\Or(N)$, $\SO(N)$, $\Sp(N)$ (i.e. the subgroup of $\GL(2N)$). This property is in fact preserved by taking direct products, but not by subgroups, and not even by quotients or central extensions \cite{Levy}. Thus the question in general needs to be re-examined for every gauge group. In particular, these results \textit{do not} cover the general case of Lie groups (for example it is known to be false for some noncompact Lie groups \cite{Sengupta}) or even compact Lie groups; for example, one could construct counterexample gauge groups of the form $G \ltimes H$, where $G$ is a finite group and $H$ a compact Lie group. In fact groups of this form have attracted interest recently \cite{Semiabelian, Completeness}. Among finite groups of interest, Wilson loops determine gauge-invariant states for the symmetric group $S_N$ \cite{Durhuus}, as well as for the dihedral group $D_4$ and the quaternion group $Q$ which we consider in Section \ref{sec:ground state wavefunction}.

As is clear from the above discussion, when we say that Wilson loops determine gauge-invariant states, we must in general consider Wilson loops in \textit{all} irreducible representations. Whether it is  sufficient to consider Wilson loops in any one (faithful) irrep is a separate question. In particular, Wilson loops in the fundamental representation are sufficient for the usual cases of $\U(N)$ and $\SU(N)$, but not (for example) for $\SO(2N)$ \cite{Durhuus}.

\subsection{Matrix elements of the Hamiltonian}\label{sec:matrix elements}

The only remaining piece in order to be able to use the holonomy basis is to compute the matrix elements of relevant operators. All operators which are gauge-invariant can be expressed in the holonomy basis. In particular, we compute the matrix elements of the Hamiltonian, which will be useful in Section \ref{sec:ground state wavefunction}.

The magnetic Hamiltonian is diagonal in the group-element basis, and is therefore also diagonal in the holonomy basis. Each (untraced) plaquette can be expressed (up to conjugation) as a product of the fundamental holonomies. Once this is done, it is easy to compute the value of the magnetic Hamiltonian on each holonomy state.

On the other hand, the electric Hamiltonian is \textit{not} diagonal. Yet its matrix elements are not hard to compute. On each link, the electric Hamiltonian is given by
\begin{equation}
    h_E = \abs{\Gamma} \mathbbm{1} - A \ , \qquad A \equiv \sum_{k \in \Gamma } L_k \ ,
\end{equation}
where we have removed the constant for ease of computation. Since $h_E$ is gauge-invariant, so is $A$ (this is also not hard to check explicitly using the properties of $\Gamma$ given in Section \ref{sec:hamiltonian formulation}). Now consider the action of $A$ on an arbitrary holonomy state, $A \ket{\{h\}}$. As we have seen from Proposition \ref{prop:one-to-one corr}, configurations with given holonomies are in one-to-one correspondence with internal gauge transformations. Therefore we can express the holonomy state as
\begin{equation}
    \ket{\{h\}} = \tfrac{1}{\sqrt{\abs{G}}^{V-1}} \sum_{\{g\} \in [h]} \ket{\{g\}} = \tfrac{1}{\sqrt{\abs{G}}^{V-1}} \sum_{\mathcal{G} \in \mathcal{G}_{\mathrm{int}}} \mathcal{G} \ket{\{h\}_{\mathrm{GF}}} \ .
\end{equation}
Since $A$ is gauge-invariant it commutes with $\mathcal{G}$, and we then see that
\begin{equation}
    A \ket{\{h\}} = \tfrac{1}{\sqrt{\abs{G}}^{V-1}} \sum_{\mathcal{G} \in \mathcal{G}_{\mathrm{int}}} \mathcal{G} A \ket{\{h\}_{\mathrm{GF}}} \ .
\end{equation}
Since applying $L_k$ to a configuration state produces another configuration state, applying $A$ to the gauge-fixed state $\ket{\{h\}_{\mathrm{GF}}}$ produces $\abs{\Gamma}$ configuration states $\ket{\{g'\}}, \ket{\{g^{\prime\prime}\}}, \ldots$ with holonomies $\{h'\}, \{h^{\prime\prime}\}, \ldots$. But then acting with all internal gauge transformations on a configuration state $\ket{\{g'\}}$ with holonomies $\{h'\}$, produces precisely the holonomy state $\ket{\{h'\}}$, i.e.
\begin{equation}
    \ket{\{h'\}} = \tfrac{1}{\sqrt{\abs{G}}^{V-1}} \sum_{\mathcal{G} \in \mathcal{G}_{\mathrm{int}}} \mathcal{G} \ket{\{g'\}} \ .
\end{equation}
As such, the action of $A$ on holonomy states takes the form
\begin{equation}
    A \ket{\{h\}} = \ket{\{h'\}} + \ket{\{h^{\prime\prime}\}} + \cdots \ .
\end{equation} 
Therefore the matrix elements of $A$ can be computed as follows:
\begin{enumerate}
    \item For each holonomy state $\ket{\{h\}}$ produce the gauge-fixed configuration $\ket{\{h\}_{\mathrm{GF}}}$.
    \item For each $k \in \Gamma$, consider the configuration $\ket{\{g'\}}=L_k \ket{\{h\}_{\mathrm{GF}}}$ and compute its holonomies $\{h'\}$.
    \item For each of these, set the matrix element $\bra{\{h'\}} A \ket{\{h\}} = 1$.
\end{enumerate}
All matrix elements which are not set to one are equal to zero. This is repeated to construct $A$ (and therefore $h_E$) for each link, and then the electric Hamiltonian is obtained by addition, $H_E = \sum_{\mathrm{links}} h_E$. Note that all operators, including $H_B$ and $H_E$, are expressed in the holonomy basis as $\abs{G}^M \times \abs{G}^M$ matrices. 

Since the holonomies are defined as products of the gauge field on up to $\mathcal{O}(V)$ links, one might have naively expected that, even on one link, the electric Hamiltonian $h_E$ would have involved up to $\mathcal{O}(V)$ terms. Instead, as we have seen, $h_E$ involves only $\mathcal{O}(1)$ terms, which is favourable.

From this description, we see that the electric Hamiltonian in the holonomy basis is guaranteed to be sparse. In particular, for each $h_E$ on each row one has $\abs{\Gamma}$ non-zero elements. Since we have $L$ links, then $H_E$ has at most $\abs{\Gamma} L$ non-zero elements per row, therefore it will overall have order $\approx \abs{\Gamma} L \abs{G}^M$ non-zero elements, i.e. the proportion of non-zero elements is an exponentially small number, $\approx \abs{\Gamma} L / \abs{G}^M$. Notably, $H_E$ is also integer-valued. 

If the group $G$ is Abelian, the matrix elements can be described more explicitly. In fact, in the Abelian case the $L_k$ operator on each link is actually gauge-invariant, and therefore maps holonomy states to holonomy states. For clarity, we denote the $L_k$ operator acting on link $l$ as $L_k \lvert_l$. As we will show in a moment, if a link $l$ occurs in the path of a fundamental holonomy more than once, then it occurs exactly twice and with opposite orientation. Therefore, since the group is Abelian, one in fact has
\begin{equation}
    L_k \lvert_l \ket{\{h\}} = \ket{\{\widetilde{h}\}} \ , \qquad \widetilde{h}_\gamma = k^{\sigma(l, \gamma)} h_\gamma  \,
\end{equation}
where $\sigma(l, \gamma) = \pm 1$ depending on whether the holonomy path $\gamma$ crosses $l$ in the positive or negative orientation, and is zero if $\gamma$ does not cross $l$. Therefore the matrix elements of $L_k$, from which those of $h_E$ can be easily deduced, are given by
\begin{equation}
    \bra{\{h'\}} L_k \lvert_l \ket{\{h\}} = \prod_{ \gamma \not\ni l} \delta(h'_\gamma, h_\gamma) \prod_{\gamma \ni l} \delta(h'_\gamma h_\gamma^{-1}, k^{\sigma(l, \gamma)}) \ ,
\end{equation}
where the first product runs over those fundamental holonomies which do not go through link $l$, and the second product goes over those that do. It remains to prove the above statement, with which we close this section.

\begin{proof}
    A path $\gamma$ defining a fundamental holonomy is of the form $\gamma = \gamma_1 \circ \widetilde{l} \circ \gamma_2$, where $\gamma_1$ and $\gamma_2$ lie entirely inside the chosen spanning tree and $\widetilde{l}$ is a link outside the tree. By construction, each link appears in $\gamma_i$ at most once. Therefore it can appear at most twice in $\gamma$. So we have to show that if a link $l$ appears in both $\gamma_1$ and $\gamma_2$, then it appears with opposite orientations. Suppose on the contrary that it appears with the \textit{same} orientation. If $l$ connects the sites $x$ and $y$ according to the orientation defined by the paths (in either the positive or negative orientation), then there are subpaths $\gamma_i' \subset \gamma_i$ (therefore lying entirely inside the maximal tree) such that $\gamma_1'$ connects the root $x_0$ to $x$ and $\gamma_2'$ connects $y$ to $x_0$. But then the closed path $\gamma_1' \circ l \circ \gamma_2'$ lies entirely within the tree. Therefore it must be the \textit{trivial} closed path. But this is impossible, since by construction neither $\gamma_1'$ nor $\gamma_2'$ contains $l$.
\end{proof}

\subsection{Topology and center symmetry}\label{sec:topology}

In this section, we discuss further structure of the Hilbert space and how it interacts with the holonomy basis. Similarly to the toric code \cite{Kitaev2003}, we discuss two kinds of topological operators, those which are diagonal in the group element basis (i.e. holonomies wrapping around non-contractible loops on the lattice) and those which are diagonal in the dual basis where the electric Hamiltonian is diagonal (these are related to center symmetry). A discussion of these issues in a similar language was given in \cite{QuantumClock} for Abelian gauge theories on a ladder. Here we extend the discussion to non-Abelian gauge theories on an arbitrary geometry.

First of all we discuss why in Yang-Mills gauge theories the electric Hamiltonian does not commute with Wilson lines or loops. Suppose that we want to fix a certain holonomy $h_i$, or even a product of holonomies. This is only possible if two separate conditions are satisfied:
\begin{enumerate}
    \item \textit{Compatibility with gauge-invariance}. As we have seen, under external gauge transformations the holonomy $h_i$ transforms as $h_i \to g h_i g^{-1}$ for $g \in G$. In order to fix this holonomy, it must be invariant under gauge transformations, i.e. $g h_i g^{-1} = h_i$ for all $g \in G$. In other words, $h_i$ must lie in the center of the gauge group $Z(G)$. The same is true for any product of holonomies, since it transforms in the same way. 
    \item \textit{Compatibility with time evolution}. An holonomy can only be consistently fixed if it cannot change value under time evolution. Since the magnetic Hamiltonian is diagonal in the holonomy basis, it does not cause transitions between states with different holonomies. The electric Hamiltonian, on the other hand, always causes transitions between different holonomy states. In particular, since the matrix elements of the electric Hamiltonian on each link are either $0$ or $1$, it suffices to show that this is true at any one link (in other words, there cannot be cancellations when summing over all links). But then consider the electric Hamiltonian on the link outside the maximal tree which defines the holonomy $h_i$; the electric Hamiltonian on that link always has non-zero matrix elements between states with different values of $h_i$, while leaving unchanged all other holonomies. Thus fixing any holonomy, or any product of holonomies, is incompatible with time-evolution according to the Hamiltonian eq.\eqref{eq:hamiltonian}.
\end{enumerate}

One may nonetheless consider other topological operators, which correspond to \textit{dual} winding. Consider a unitary operator of the form
\begin{equation}
    \label{eq:generic thooft}
    U(\{g\}) = \bigotimes_{l \in \mathrm{links}} L_{g_l}  \ ,
\end{equation}
where $\{g\}$ is a configuration, i.e. an assignment of a group element $g_l$ to each lattice link $l$. Thus in principle $U(\{g\})$ acts on all lattice links, but we will soon impose further conditions which will restrict its form. Eq.\eqref{eq:generic thooft} generalizes an operator for the Abelian case in \cite{QuantumClock}. As we will soon see, it is irrelevant whether one chooses left or right translations to define $U$. Again $U$ is only useful if it is compatible with the gauge symmetry (i.e. it must be gauge-invariant) and time evolution (i.e. it must be a symmetry of the Hamiltonian).

Consider an arbitrary gauge transformation $\mathcal{G}$, as defined in eq.\eqref{eq:gauge transformation}. Then one can check that 
\begin{align}
    \mathcal{G} U(\{h\}) \mathcal{G}^{-1} &= \bigotimes_{l=\expval{xy} \in \mathrm{links}} L_{g_x} R_{g_y} L_{h_l} R_{g_y^{-1}} L_{g_x^{-1}} = \nonumber\\
    &=\bigotimes_{l=\expval{xy} \in \mathrm{links}} L_{g_x h_l g_x^{-1}} \ .
\end{align}
Therefore gauge-invariance requires $g_x g_l g_x^{-1} = g_l$ for all $g_x$. Thus $g_l$ must lie in the center of the gauge group $G$. In other words, the operator $U(\{g\})$ is gauge-invariant if and only if $\{g\}$ is a configuration made entirely of central elements. Note that if $g$ is central, then $R_g = L_{g^{-1}}$ so restricting to left translations does not lead to loss of generality.

Now consider time evolution. Using the invariance of $\Gamma$ under conjugation, it is not hard to see that, for arbitrary $g$ and on any link,
\begin{equation}
    L_g h_E = h_E L_g \ .
\end{equation}
Therefore $U(\{g\})$ always commutes with the electric Hamiltonian. Note that this is the \enquote{dual} statement to the fact that the magnetic Hamiltonian preserves the holonomies. 

We then need to consider conditions under which $U(\{g\})$ commutes with the magnetic Hamiltonian. Until now we could work on an arbitrary graph geometry, but the magnetic Hamiltonian requires a bit more structure, so we specialize to a hypercubic lattice. Nonetheless the generic construction worked out in this section can be used to find appropriate operators on arbitrary geometries. As we have seen above, the configuration $\{g\}$ is required to be central for gauge-invariance. Then since $U(\{g\})$ is gauge-invariant, it maps holonomy states to holonomy states. Therefore the fact that the magnetic Hamiltonian commutes with $U(\{g\})$ is equivalent to stating that $H_B$ takes the same value on the configuration states $\ket{\{g'\}}$ and $U(\{g\})\ket{\{g'\}}$ for arbitrary $\{g'\}$. Since the $g_l$s are central, plaquettes transform as
\begin{equation}
    \label{eq:plaquette preservation condition}
    g'_1 g'_2 g_3^{\prime -1} g_4^{\prime -1} \to g_1 g_2 g_3^{-1} g_4^{-1} g'_1 g'_2 g_3^{\prime -1} g_4^{\prime -1} \ .
\end{equation}
As we show in Appendix \ref{sec:properties faithful irreps} the magnetic Hamiltonian on each plaquette, $h_B = - 2 \Re \chi$ has a unique minimum at the identity, since $\chi$ is assumed to be the character of a faithful representation. Since eq.\eqref{eq:plaquette preservation condition} must hold for all configurations, it must in particular hold for configurations $\{g'\}$ where the plaquette variable $g'_1 g'_2 g_3^{\prime -1} g_4^{\prime -1}$ is equal to the identity. Thus it is a necessary and sufficient condition that $g_1 g_2 g_3^{-1} g_4^{-1}=1$ for all plaquettes. In other words, the configuration $\{g\}$ must be a flat configuration living in the center of the gauge group $G$. Note that this does \textit{not} mean in general that $\{g\}$ is gauge-equivalent to the identity.

Up until this point, we have therefore constructed a large number of gauge-invariant symmetry operators $U(\{g\})$. But, as we will now show, most of these act in the same way in the gauge-invariant subspace. Suppose in fact that we construct another operator $U(\{\widetilde{g}\})$ which is related to $U(\{g\})$ by a gauge transformation of the $g_l$s, i.e. $\widetilde{g}_{\expval{xy}} = k_x g_{\expval{xy}} k_y^{-1}$ for some $k_x \in G$. Note that a gauge transformation of $g_l$s does \textit{not} correspond to a gauge transformation of $U$. Gauge invariance requires that both $\{g\}$ and $\{\widetilde{g}\}$ are central; it is not hard to show that one may then choose all $k_x$ to be central also. Then one finds that
\begin{equation}
    U(\{\widetilde{g}\}) = \mathcal{G} U(\{g\}) \ ,
\end{equation}
where $\mathcal{G}$ is a gauge transformation of the usual form (i.e. eq.\eqref{eq:gauge transformation}) with group elements $k_x$. In other words, performing a gauge transformation of $\{g\}$ leads to $U$ operators which coincide on the physical subspace and should therefore be viewed as physically equivalent. One can therefore perform gauge transformations of the $\{g\}$ to reduce them to the minimal amount of degrees of freedom. 

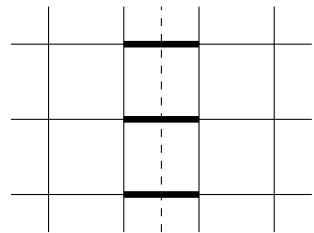
\begin{figure}
    \centering
    \begin{tikzpicture}
        \draw[step=1.0,black] (0.5,0.5) grid (4.5,3.5);
        \draw[dashed] (2.5,0.5) -- (2.5,3.5);
        \draw[line width=2.5pt] (2,1) -- (3,1);
        \draw[line width=2.5pt] (2,2) -- (3,2);
        \draw[line width=2.5pt] (2,3) -- (3,3);
    \end{tikzpicture}
    \caption{A square lattice. One of the two 't Hooft operators is supported on the bold links, i.e. those crossing the dashed line. This 't Hooft operator thus multiplies by a central element any loop winding around the horizontal direction.}
    \label{fig:thooft operator}
\end{figure}

For example, on a $d-$dimensional periodic hypercubic lattice the independent $U$ operators are the 't Hooft operators responsible for center symmetry. In this case, a flat central configuration $\{g\}$ can be gauge-transformed to one which is the identity everywhere except on the links which pierce a $d-1$ dimensional hypersurface. On these links, it is equal to a constant central element. The 't Hooft operator then takes the form \cite{HarlowOoguri}
\begin{equation}\label{eq:thooft operator}
    U_{S}(g) = \bigotimes_{l \in S} L_{g} \ ,
\end{equation}
where $g$ is central and $S$ is the set of links which pierces a $d-1$ dimensional surface. Since $U$ operators related by a gauge transformation of the $g_l$s are physical equivalent, the location of the surface $S$ is immaterial. Therefore the $U$ operators are topological. On a $d$-dimensional periodic hypercubic lattice, one therefore has $d$ independent such operators which cannot be deformed into each other. An example of such an operator on the square lattice is given in Fig.~\ref{fig:thooft operator}. Here the surface $S$ is one dimensional, i.e. it is a line.

\begin{table*}[t]
    \begin{center}
    \setlength{\tabcolsep}{4pt}
    \begin{tabular}{cccccccccc}
    \toprule
    Size & $V$ & $L$ & $M$ & $\dim \Htot$ & $\dim \Hhol$ & $\dim \Hphys$ & \texttt{time}$(H_E)$ & \texttt{size}$(H_E)$ & \texttt{sparsity}$(H_E)$ \\
    \midrule
    $2\times 2$ & $4$ & $8$ & $5$ & $2^{24}$ & $2^{15}$ & $8960 \approx 2^{13}$ & $0.1 \, \mathrm{s}$ & $13 \, \mathrm{MB}$ & $0.15\%$\\
    $2\times 3$ & $6$ & $12$ & $7$ & $2^{36}$ & $2^{21}$ & $536576 \approx 2^{19}$ & $10\, \mathrm{s}$ & $1.2 \, \mathrm{GB}$ & $0.0035\%$\\
    \bottomrule
    \end{tabular}
    \end{center}
    
    \caption{Parameters for the two lattices used in the simulations for the two gauge groups $D_4$ and $Q$. All quantities happen to be either approximately or exactly the same for both groups. The labels \texttt{time}$(H_E)$, \texttt{size}$(H_E)$ and \texttt{sparsity}$(H_E)$ refer respectively to the time taken to construct $H_E$ (on a common laptop), its size and the proportion of its non-zero elements.}
    \label{tab:parameters}
\end{table*}

In the confining phase of Yang-Mills theory, one expects center symmetry to be unbroken. One could then further reduce the size of the relevant subspace by considering only states which are center-invariant. In fact it is rather simple to construct such a projector, and one can even compute the size of the relevant gauge-invariant, center-invariant subspace following an argument similar to Theorem \ref{theo:Hphys size}. However, this goes against the philosophy of the holonomy states, whose main advantage is precisely the fact that the description of their Hilbert space does not require Clebsch-Gordan coefficients.

\section{Ground state wavefunction}\label{sec:ground state wavefunction}

The determination of the ground state wavefunction of Yang-Mills theory is not only a problem of theoretical interest, but also relevant in the quantum simulation setting where some quantum algorithms require a sufficiently good approximation to the ground state as a starting point.

The ground states for the electric and magnetic Hamiltonians can be easily described in the holonomy basis. While the electric Hamiltonian can have ground state degeneracy \cite{MPE}, we do not consider this case here. On each link, the ground state of the electric Hamiltonian is then given by an equal superposition of all group elements. Tensoring across all links, we find the electric ground state
\begin{equation}
    \ket{0_E} = \frac{1}{\sqrt{\abs{G}}^{M}} \sum_{\{h\}} \ket{\{h\}} \ .
\end{equation}
Since it corresponds to the constant wavefunction it is gauge-invariant and it is the maximal entropy state in the holonomy basis. On the other hand the magnetic Hamiltonian generally has several ground states. Usually one chooses $h_B=-2\Re\chi$ where $\chi$ is the character of a faithful representation, so that it is minimal on the identity element only (this we prove in Appendix \ref{sec:properties faithful irreps}). Then the ground states are those with all plaquettes equal to the identity. The simplest gauge-invariant ground state is rather easy to write in the holonomy basis, i.e.
\begin{equation}\label{eq:magnetic ground state}
    \ket{0_B} = \ket{1,1,\ldots, 1} \ .
\end{equation}
This is gauge invariant because the identity is a singlet under conjugation. This is also a state of minimal entropy in the holonomy basis. Note that the preparation of both $\ket{0_E}$ and $\ket{0_B}$ on a quantum device is standard. 

In the group element basis, it has been argued \cite{Feynman:1981ss, Greensite79, Greensite1980, fieldstrength} that the confining ground state of lattice Yang-Mills theory in $2+1$ dimensions (here we take a periodic square lattice) can be approximated by a trial wavefunction of the form 
\begin{equation}
    \label{eq:ground state ansatz}
    \psi_{\alpha}(\{g\}) = \exp{\pqty{-\alpha \sum_\square \Re \chi(g_{\square}) }} \ ,
\end{equation}
where $\alpha$ is a variational parameter and $\chi$ is the group character which appears in the magnetic Hamiltonian (and, for relativistic theories, also in $h_E$ via eq.\eqref{eq:relativistic gamma}). The ansatz eq.\eqref{eq:ground state ansatz} interpolates between the electric ground state ($\alpha = 0$) and the magnetic ground state ($\alpha=-\infty$). For each value of the coupling $\lambda$ of the Hamiltonian $H$ (recall eq.\eqref{eq:hamiltonian reduced lambda}), $\alpha$ is determined by numerically minimizing the average energy per link
\begin{equation}\label{eq:variational H expval}
    E(\alpha) \equiv \frac{1}{L} \frac{\bra{\psi_\alpha} H \ket{\psi_\alpha} }{ \expval{\psi_\alpha | \psi_\alpha} } \ .
\end{equation}
Due to the peculiarities of two space dimensions, this minimization is rather easy and can be performed directly in infinite volume. The details are described in Appendix \ref{sec:variational minimization}.

The trial wavefunction in eq.\eqref{eq:ground state ansatz} has been used in several variational calculations for Lie group gauge theories both in $2+1$ and higher dimensions \cite{Arisue, Arisue2, Dagotto, Dabringhaus, Guo, McIntosh2002, McIntosh2002_2, ImprovingHam}. It can also be systematically improved by adding larger Wilson loops, for example a $2 \times 1$ rectangular loop. Other approaches have been developed towards variational vacuum wavefunctionals,  but they are generally more complicated and often not easily applicable to finite gauge groups. The wavefunction eq.\eqref{eq:ground state ansatz} is particularly appealing due to its simplicity. Note that our chosen trial wavefunction is based on the idea of an expansion in terms of Wilson loops, and it is possible that it might not be applicable to gauge theories where Wilson loops do not determine the gauge invariant states (as discussed in Section \ref{sec:wilson loops}). However the groups considered here do not suffer from this problem.

\begin{figure*}
    \centering
    \subfloat{\includegraphics[width=0.48\textwidth]{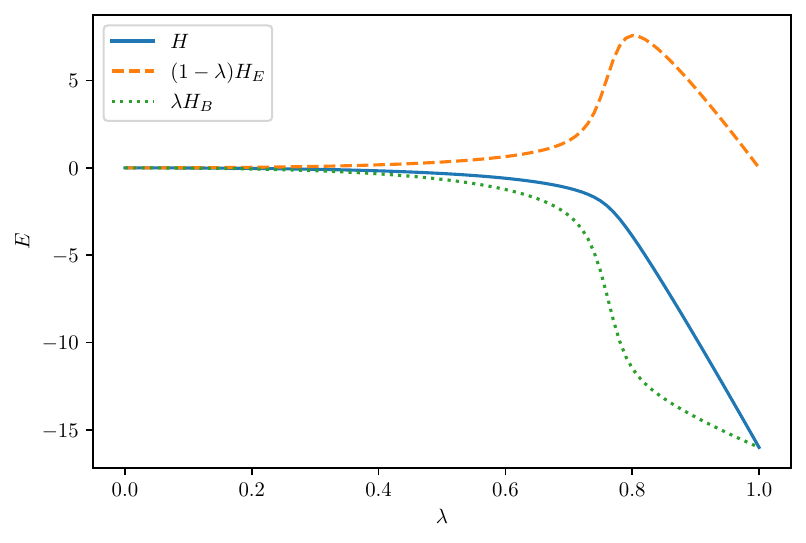}}
    \quad
    \subfloat{\includegraphics[width=0.48\textwidth]{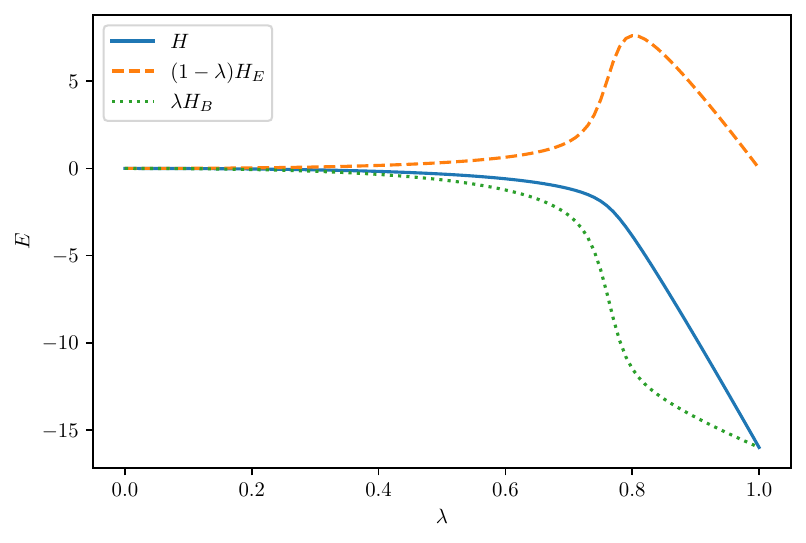}}
    \quad
    \caption{Ground state energy on a $2 \times 2$ lattice for $D_4$ (\textit{left}) and $Q$ (\textit{right}), as well as expectation value of $H_E$ and $H_B$ in the ground state. The two plots are almost indistinguishable.}
    \label{fig:ground state energies} 
\end{figure*}

In \cite{TestingProposals, Greensite1987, Greensite1989, Arisue3}, Monte Carlo methods were employed to compare different proposals for ground state wavefunctions (all closely related to eq.\eqref{eq:ground state ansatz}) on some sets of configurations for non-Abelian Lie groups. Here we simply compute the overlap (i.e. the fidelity) between the trial wavefunction and the exact finite-volume ground state. Of course, due to the exponential size of the Hilbert space, we are limited to small lattices.

In what follows, we consider two different finite gauge groups of order $8$. These are the dihedral group $D_4$ and the quaternion group $Q$, which are subgroups of $\Or(2)$ and $\SU(2)$ respectively. In both cases, we choose the character of their unique faithful irrep for the magnetic Hamiltonian, and then set the electric Hamiltonian according to eq.s~\eqref{eq:electric hamiltonian link} and \eqref{eq:relativistic gamma}. Using holonomy states, we construct their Hamiltonian on two small lattices, $2\times 2$ and $2\times 3$ with periodic boundary conditions. Note that, using spin network states, only a $2\times 2$ lattice was possible \cite{MPE}. The relevant parameters for the two cases are given in Table \ref{tab:parameters}. Since $D_4$ and $Q$ (accidentally) share the same character table, they also share the same physical subspace dimension (on any graph). Moreover, as we will soon see, despite the fact that their Hamiltonians are \textit{not} equal, the results for the two groups are very similar.

Given each lattice, we construct the Hamiltonian in the holonomy basis following the recipe provided in Section \ref{sec:matrix elements}. We can then determine some of the lowest eigenstates and compute several quantities of interest. In particular, we can compute the overlap of the true ground state with the trial wavefunction eq.\eqref{eq:ground state ansatz}. Our code was implemented in Python compiled with Numba \cite{lam2015numba}; the Hamiltonian (and all other matrices) were implemented in sparse matrix format with SciPy \cite{2020SciPy-NMeth}. In terms of constructing operators, the electric Hamiltonian was by far the slowest and we have reported the time required to construct it (on a common laptop) in Table \ref{tab:parameters}. This is of course not a precise benchmark but serves to give an idea of how efficient the code can be in practice. Among the various matrices which we need to store, $H_E$ is by far the largest in terms of memory use, and again we list its size in Table \ref{tab:parameters}.

\begin{figure*}
    \centering
    \subfloat{\includegraphics[width=0.48\textwidth]{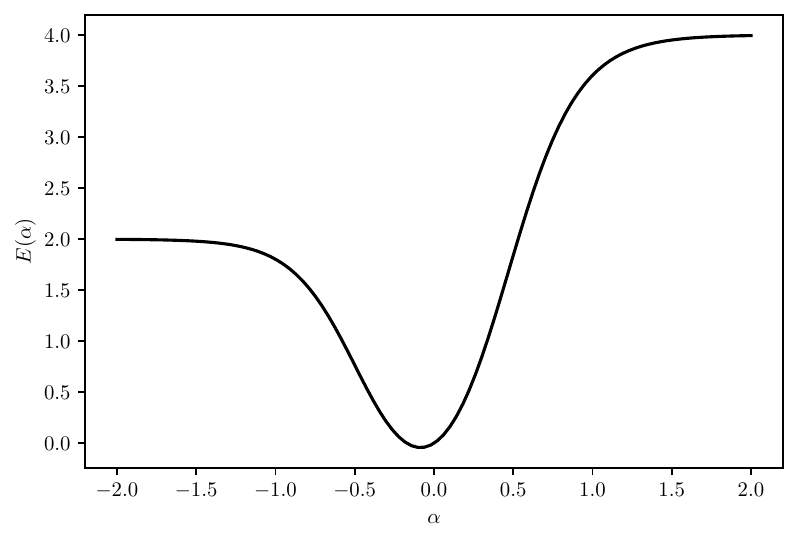}}
    \quad
    \subfloat{\includegraphics[width=0.48\textwidth]{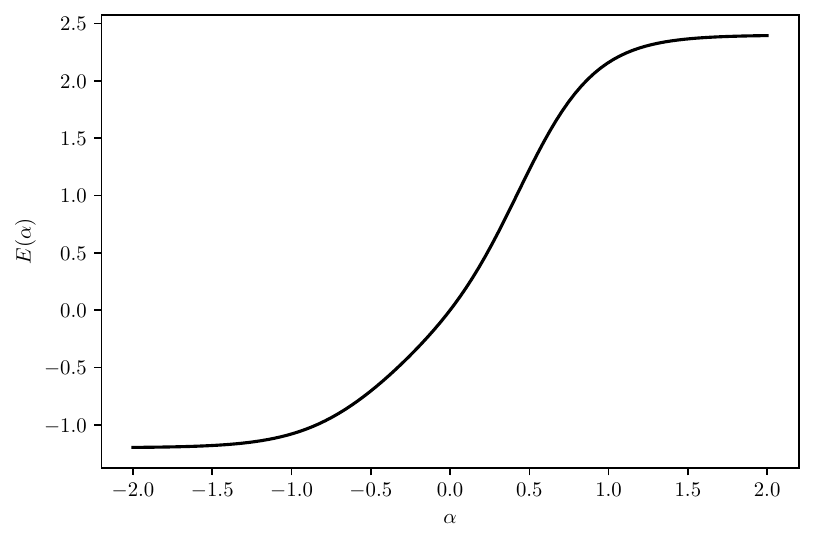}}
    \quad
    \caption{Variational energy $E(\alpha)$ of eq.\eqref{eq:variational H expval} for $D_4$ gauge theory as a function of $\alpha$ for two values of $\lambda$, i.e. $\lambda=0.5$ (\textit{left}) and $\lambda=0.9$ (\textit{right}). For small $\lambda$, $E(\alpha)$ achieves a minimum, while for large $\lambda$ the energy asymptotes to $\alpha = -\infty$. }
    \label{fig:variational energies} 
\end{figure*}

An important consideration is that one still needs to impose invariance under the external gauge transformation. Generally speaking, the holonomy space is a constrained Hilbert space much like the total Hilbert space. Thus they should be treated similarly. On a quantum device, one would map the holonomy Hilbert space $\Hhol$ directly onto the quantum register. On a classical device, a gauge-invariant basis could be constructed by numerically solving the equation $P\ket{\psi} = \ket{\psi}$, where $P$ is the projector onto the gauge-invariant subspace in the holonomy basis of eq.\eqref{eq:gauge projector holonomies}. This is already advantageous compared to performing the same operation in the total Hilbert space, as the memory requirements for storing the states are substantially reduced in the holonomy basis.

However, in our case, it is sufficient to diagonalize the Hamiltonian directly in the holonomy basis. One could then apply the projector $P$, which is a cheap operation, and filter out identical states. But even this is not required in most cases. In fact suppose that $\ket{\psi}$ is a non-degenerate energy eigenstate of the Hamiltonian $H$; since the electric and magnetic Hamiltonians do not commute, this is the typical situation. Since $P$ commutes with $H$, then if $P\ket{\psi}$ is not zero, it is an energy eigenstate with the same energy and therefore $P\ket{\psi} = \ket{\psi}$ (since the state is non-degenerate and the eigenvalues of $P$ are either $0$ or $1$). In other words, in this case states are either gauge-invariant or not, and no superpositions occur. It is therefore quite easy to check for gauge-invariance. In particular, in all our simulations, the finite-volume ground state of the Hamiltonian was always found to be unique and gauge-invariant.

We now discuss the results of our numerical simulations. In $2+1$ dimensions, gauge theories with non-Abelian finite groups generally have a two-phase structure, with a confining phase for small $\lambda$ (where the electric Hamiltonian dominates) and a deconfined phase for large $\lambda$ (where the magnetic Hamiltonian dominates), separated by a first-order transition \cite{Rebbi,BhanotCreutz,SU3Subgroups,SU2Subgroups,ZNgeneralizedAction}. 

Fig.~\eqref{fig:ground state energies} shows the ground state energy, as well as the expectation values of the electric and magnetic Hamiltonians in the ground state. It has been previously shown that the location of the phase transition may be reasonably identified with the point of sharpest change of the expectation value of the electric Hamiltonian \cite{MPE}. With this prescription, we find the pseudocritical coupling $\lambda_c^{(2\times 2)} = 0.76(1)$ for both groups on the $2 \times 2$ lattice (for $D_4$, in agreement with \cite{MPE}). On the $2 \times 3$ lattice the pseudocritical coupling increases slightly to $\lambda_c^{(2\times 3)} = 0.77(1)$, again for both groups.

\begin{figure}
    \centering
    \includegraphics[width=0.48\textwidth]{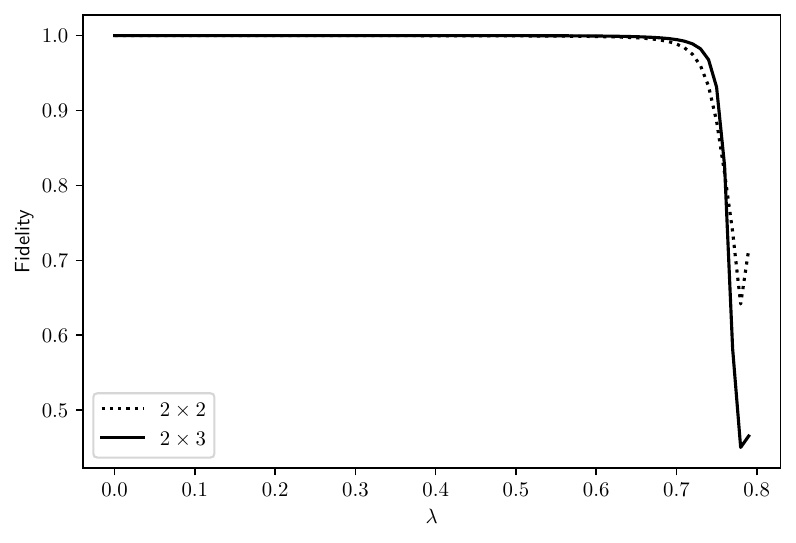}
    \caption{Fidelity between the exact finite volume ground state and the the ground state ansatz eq.\eqref{eq:ground state ansatz} for $D_4$ gauge theory on a $2 \times 2$ and $2 \times 3$ lattice, for $\lambda \in [0, 0.79]$. The picture for $Q$ is quite similar. The agreement is excellent for small $\lambda$ but drops substantially near the phase transition, especially for the larger lattice. }
    \label{fig:overlaps}
\end{figure}

On the other hand, Fig.~\ref{fig:variational energies} shows the variational energy $E(\alpha)$ defined in eq.\eqref{eq:variational H expval} for the group $D_4$ in two cases, $\lambda = 0.5$ and $\lambda = 0.9$. In particular, $E(\alpha)$ achieves a minimum for small $\lambda$, but not for large $\lambda$, where the energy asymptotes to its minimum at $\alpha = -\infty$. Therefore the minimization procedure is well-defined only for $\lambda$ sufficiently small. This should not be thought of as a limitation of the ground state ansatz eq.\eqref{eq:ground state ansatz}, but rather as a feature. Since the trial wavefunction defines a confining ground state, it is not expected to be valid in the deconfined phase. Thus the failure of the minimization should be taken as a \textit{prediction} for the location of the phase transition. A similar situation was encountered in \cite{HeysStump}. With this definition, we locate the phase transition for both groups at $\lambda_c = 0.79(1)$. Note that this value obtained via the infinite-volume trial wavefunction is completely independent from the one obtained from the exact finite-volume numerical data, yet quite close. 

In the confining phase, we can compare the exact finite-volume ground state with the ground state ansatz eq. \eqref{eq:ground state ansatz}. Given two normalized states $\ket{\psi_1}$ and $\ket{\psi_2}$, a natural measure of their similarity is their \textit{fidelity}, i.e. the absolute value squared of their overlap $\abs{\expval{\psi_1 | \psi_2}}^2$. In Fig.~\ref{fig:overlaps} we plot the fidelity between the exact finite-volume ground state and the trial wavefunction in the confining phase. In particular, we see that the trial wavefunction provides an almost perfect approximation to the exact ground state for much of the confining phase, but near the phase transition the approximation is less good. Moreover, the fidelity decreases more sharply on the larger $2 \times 3$ volume.

Overall, we find that the ansatz eq.\eqref{eq:ground state ansatz} provides a good approximation for the ground state of the two theories, especially in that it is able to predict the existence of the phase transition, as well as its location with reasonable accuracy. However, it does not fully capture the correlations near the phase transition, which is the most interesting location. Further, extrapolating from the trend in Fig.~\ref{fig:overlaps}, one expects the fidelity near the transition to further decrease on larger volumes. It would be interesting to study the effect of adding larger loops to the trial wavefunction; we would expect them to improve its accuracy. 

\begin{figure}
    \centering
    \includegraphics[width=0.48\textwidth]{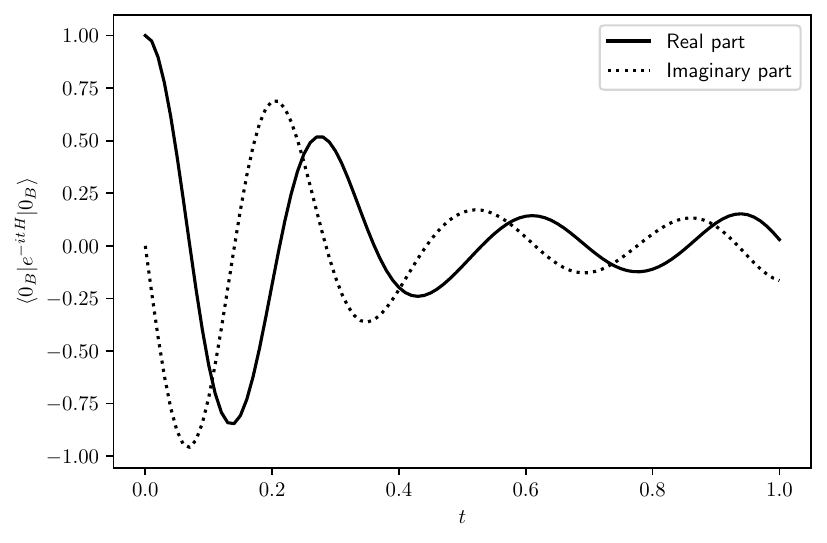}
    \caption{Expectation value of the time-evolution operator with the full Hamiltonian $H$ at $\lambda = 0.4$ in the magnetic ground state $\ket{0_B}$ (eq.\eqref{eq:magnetic ground state}), for $D_4$ gauge theory on the $2 \times 2$ lattice.}
    \label{fig:time evolution}
\end{figure}

We also examined what happens to the 't Hooft operators defined in Section \ref{sec:topology}. Both $D_4$ and $Q$ have center isomorphic to $\Z_2$. Therefore on a periodic square lattice, one can construct two independent 't Hooft operators according to eq.\eqref{eq:thooft operator}. Numerically, we found in all cases that the finite-volume ground state (but not excited states) is invariant under both operators for all values of $\lambda$. Finally, in Fig.~\ref{fig:time evolution} a quench for $D_4$ on the $2 \times 2$ lattice where the magnetic ground state is time-evolved using the full Hamiltonian $H$ at $\lambda=0.4$. The results for $Q$ are similar.

Calculations of the kind presented here could be used as benchmarks for quantum simulators. As is clear from the trend in Table \ref{tab:parameters}, using our current implementation it is not possible to extend the calculation to a $3 \times 3$ lattice due to the large memory requirements. One way to substantially reduce the size of the Hilbert space, and therefore memory requirements, is to give up periodicity, which may even be more appropriate for comparison with quantum simulators. Nonetheless, with our current implementation it should be possible to exactly diagonalize non-Abelian groups of order up to approximately $16$ on a $2 \times 2$ periodic lattice.

\section{Conclusions}

In this work, we have discussed how to represent the physical, gauge-invariant Hilbert space of pure lattice gauge theories in terms of a class of \enquote{holonomy states}. Based on this construction, we discussed the Wilson loop representation of the gauge-invariant space as well as the topological structure of the Hilbert space. On small lattices, we could use this construction to exactly diagonalize the Hamiltonian. In particular, we computed the exact overlap between the finite-volume ground state and a conjectured ansatz and discussed to what extent they agree.

As we have seen, the holonomy basis may be useful to study small systems in classical simulations, and possibly also in current-era quantum devices. It would be interesting to understand whether the holonomy basis can be efficiently implemented on a quantum device. The crucial difficulty here is the construction of the electric Hamiltonian. It would also be interesting to study how to efficiently prepare the ground state wavefunction eq.\eqref{eq:ground state ansatz} on a quantum device.

As mentioned in the beginning, whether working in the total or physical Hilbert space is more advantageous may depend on what quantum device is used for the simulation. It is currently unclear what kind of hardware architecture might prevail in the long term and therefore allow quantum simulation of gauge theories at scale. It is also possible that multiple architectures could co-exist for different purposes, much like CPUs and GPUs for classical computing. On an architecture where qubits are cheap, it is unlikely that working in terms of physical variables is helpful. On the other hand, on an architecture where qubits are expensive but, for example, gates can be accurately applied, then it might be useful to work in a physical basis.

It would be interesting to further explore other properties of the holonomy basis. For example, it is unclear to what extent the properties of the holonomy states depend on the choice of maximal tree. Moreover, one may even think of applying various truncation methods directly in the holonomy basis, once most of the gauge symmetry has been removed.

The most natural next step is the inclusion of matter fields. Developing methods to compute the dimension of the physical subspace is particularly interesting in this case. For example, in the presence of a gauge anomaly there are no gauge invariant states \cite{WittenSU(2)}. 

\section*{Acknowledgments}

The author thanks G. Kanwar, S.Pradhan, U.-J. Wiese and E. Ercolessi for useful discussions and comments. The research leading to this work has received funding from the Schweizerischer Nationalfonds (grant agreement number 200020\_200424).

\appendix

\section{Properties of faithful irreps}\label{sec:properties faithful irreps}

In Section \ref{sec:hamiltonian formulation} we defined the magnetic Hamiltonian on each plaquette to be the function $h_B \equiv -2 \Re \chi$, where $\chi$ is the character of a faithful representation of the gauge group. Several times we have used the result that $h_B$ has a unique minimum at the identity element, a result which we now prove. The results in this section are valid for finite-dimensional representations of both finite groups as well as compact Lie groups. First we show a related result:

\begin{proposition}\label{prop:chi inequality}
    Let $\rho$ be a representation of $G$ and $\chi$ its character. Then for all $g \in G$, $\abs{\chi(g)} \leq \chi(1)$ with equality if and only if $\rho(g)$ is proportional to the identity matrix.
\end{proposition}

\begin{proof}
    Since $G$ is either finite or compact Lie its finite-dimensional representations may be chosen to be unitary \cite{Serre, KnappLieGroups}. Let $\chi = \tr \rho$, where $\rho$ is a unitary representation. Then $\rho(g)$ is a unitary matrix, and as a result, its eigenvalues $\lambda_i$ are all unit-norm, i.e. $\abs{\lambda_i}=1$. Then by the triangle inequality
    \begin{equation}
        \abs{\chi(g)} = \abs{\sum_i \lambda_i } \leq \sum_i \abs{ \lambda_i } = \dim\rho = \chi(1) \ ,
    \end{equation}
    with equality if and only if all the $\lambda_i$ are equal, in which case $\rho(g)$ is proportional to the identity matrix.
\end{proof}

Now we refine the previous statement of equality:

\begin{proposition}\label{prop:chi equality}
    If $\chi(g) = \chi(1)$, then $\rho(g) = \mathbbm{1}$ is the identity matrix.
\end{proposition}

\begin{proof}
    Since $\chi(g) = \chi(1)$, then also $\abs{\chi(g)} = \chi(1)$ and therefore by the previous Proposition, $\rho(g) = \lambda \mathbbm{1}$ for some $\lambda$ with $\abs{\lambda}=1$. But then
    \begin{equation}
        \chi(g) = \lambda \tr(\mathbbm{1}) = \lambda \dim{\rho} = \lambda \chi(1) \ .
    \end{equation}
    Therefore by the hypothesis $\lambda=1$ and $\rho(g)$ is the identity matrix.
\end{proof}

Now we finally prove the required statement:

\begin{proposition}
    If $\chi$ is the character of a faithful representation, then $\Re\chi(g)$ has a unique maximum at the identity element.
\end{proposition}

\begin{proof}
    Using Prop. \ref{prop:chi inequality}, we see that
    \begin{equation}
        \Re\chi(g) \leq \abs{\chi(g)} \leq \chi(1) = \Re \chi(1) \ .
    \end{equation}
    Therefore $\Re\chi(g)$ achieves its maximum at the identity element. Now suppose that $\Re\chi(g) = \Re \chi(1)$. By the above inequality, this implies that $\Re\chi(g) = \abs{\chi(g)}$, so $\chi(g)$ is real and positive. Therefore we also have $\chi(g) = \chi(1)$, which, by Prop. \ref{prop:chi equality}, implies that $\rho(g)$ is the identity matrix (where $\rho$ is the representation with character $\chi$). But since $\rho$ is faithful by assumption, then $g=1$ is the identity, proving uniqueness.
\end{proof}

\section{Variational minimization}\label{sec:variational minimization}

As remarked in Section \ref{sec:ground state wavefunction}, in order to determine the variational parameter $\alpha$ of the ground state wavefunction eq.\eqref{eq:ground state ansatz} one numerically minimizes the expectation value of the Hamiltonian eq.\eqref{eq:variational H expval}. The minimization can be performed directly in infinite volume and is quite simple due to the peculiarities of two space dimensions.

For ease of reference we define the following functions which will appear in what follows,
\begin{align}
    F_1(\alpha) &\equiv \sum_{g \in G} \exp{\pqty{-2\alpha \Re \chi(g)}} \ ,\\
    F_2(\alpha) &\equiv \sum_{g \in G} \exp{\pqty{-2\alpha \Re \chi(g)}} \Re \chi(g) \ , \\
    F_3(\alpha) &\equiv \sum_{g_1, g_2 \in G} \sum_{k \in \Gamma} \exp \big[-\alpha \Re \big( \chi(g_1) + \chi(g_2) + \\
    &\qquad\qquad\qquad\qquad +\chi(g_1 k) + \chi(g_2 k^{-1})  \big) \big] \ . \nonumber
\end{align}
The key ingredient on a two-dimensional infinite square lattice is the change of variables from links to plaquettes,
\begin{equation}
    \frac{1}{\abs{G}^L} \sum_{\{g\}} \longrightarrow \frac{1}{\abs{G}^K} \sum_{\{g_\square\}} \ ,
\end{equation}
where $K=L/2$ is the number of plaquettes. This is a classic result; in fact the change of variables from links to plaquettes is a lattice Bianchi identity \cite{fieldstrength, fieldstrengthlattice} which is trivial in two dimensions \cite{ImprovingHam, Ligterink}. From a more modern perspective, this can also be obtained from the more general change of variable formula for the holonomies eq.\eqref{eq:change of variable formula} by choosing an appropriate maximal tree for the infinite square lattice and then performing an appropriate change of variables from holonomies to plaquettes.

We can then compute first of all the normalization of the trial state,
\begin{align}
    \expval{\psi_\alpha | \psi_\alpha} &= \sum_{\{g\}} \exp{\pqty{-2 \alpha \sum_\square \Re \chi(g_{\square}) }}= \nonumber \\
    &=\frac{\abs{G}^L}{\abs{G}^K} \sum_{\{g_\square\}} \exp{\pqty{-2 \alpha \sum_\square \Re \chi(g_{\square}) }} =\\
    &=\frac{\abs{G}^L}{\abs{G}^K} F_1(\alpha)^K \ , \nonumber
\end{align}
since we can now sum each plaquette independently. A very similar calculation leads to
\begin{equation}
    \bra{\psi_\alpha} H_B \ket{\psi_\alpha} = -2 K \frac{\abs{G}^L}{\abs{G}^K} F_1(\alpha)^{K-1} F_2(\alpha) \ .
\end{equation}
For the electric Hamiltonian, one needs to consider the two plaquettes hinging on each link and then one finds,
\begin{equation}
    \bra{\psi_\alpha} h_E \ket{\psi_\alpha} = \abs{\Gamma} \expval{\psi | \psi} - \frac{\abs{G}^L}{\abs{G}^K} F_1(\alpha)^{K-2} F_3(\alpha) \ .
\end{equation}
Therefore overall the target function to minimize is
\begin{equation}
    E(\alpha) \equiv \frac{1}{L} \frac{\bra{\psi_\alpha} H \ket{\psi_\alpha} }{ \expval{\psi_\alpha | \psi_\alpha} } = (1-\lambda) \bqty{\abs{\Gamma}-\frac{F_3(\alpha)}{F_1(\alpha)^2}} - \lambda \frac{F_2(\alpha)}{F_1(\alpha)} \ .
\end{equation}
Since the $F$ functions are rather simple to compute numerically, for each $\lambda$ it is easy to minimize $E$ in order to find the optimal $\alpha$.

\bibliography{biblio}

\end{document}